\documentclass[runningheads, envcountsect]{llncs}
\usepackage[T1]{fontenc}

\usepackage{graphicx,algorithm,algorithmic,clrscode3e,tikz,bm,amssymb,amsmath,empheq,enumitem,graphicx,color,footmisc,appendix,fullpage,soul}
\def\emptp  {\textit{$\hat{mty}_{pr}$}}
\def\empts  {\textit{$\hat{mty}_{ts}$}}

\spnewtheorem{Lemma}[thm]{Lemma}{\bfseries}{\itshape}
\spnewtheorem{Corollary}[thm]{Corollary}{\bfseries}{\itshape}
\spnewtheorem{Figure}{Figure}[section]{\bfseries}{\itshape}
\spnewtheorem{Equation}{Equation}[section]{\bfseries}{\itshape}

\usepackage{enumitem}
\newlist{Property}{enumerate}{2}
\setlist[Property]{label=Property \arabic*., font=\textbf, itemindent=*}

\let\llncssubparagraph\subparagraph
\let\subparagraph\paragraph
\usepackage[compact]{titlesec}
\let\subparagraph\llncssubparagraph

\usepackage{titlesec}

\titleformat{\paragraph}
{\normalfont\normalsize\bfseries}{\theparagraph}{1em}{}
\titlespacing*{\paragraph}
{0pt}{3.25ex plus 1ex minus .2ex}{1.5ex plus .2ex}

\newcommand{\setof}[1]{\{{#1}\}}

 \makeatletter
\def\@fnsymbol#1{\ensuremath{\ifcase#1\or \dagger\or *\or *\or
		\mathsection\or \mathparagraph\or \|\or **\or \dagger\dagger
		\or \ddagger\ddagger \else\@ctrerr\fi}}
\makeatother

\begin{document}
\title{Optimal Algorithm for Paired-Domination in\\ Distance-Hereditary Graphs\thanks{A preliminary version of this work was presented at COCOON 2022. This work is partially supported by the \newline \mbox{} \hspace{11.7pt} National Science and Technology Council under the Grants No. MOST 109-2221-E-019-050-MY2.}}
\titlerunning{Optimal Algorithm for Paired-Domination in Distance-Hereditary Graphs}
%
%
\author{Ta-Yu Mu\inst{1} \and
	Ching-Chi Lin\inst{2}\thanks{Corresponding author. \newline \mbox{} \hspace{11.7pt} E-mail address: lincc@mail.ntou.edu.tw (C.-C. Lin).}}

\authorrunning{T.-Y. Mu and C.-C. Lin}
%
\institute{National Taiwan University, Taipei 10617, Taiwan\\
	\email{f08922132@ntu.edu.tw}\and
	National Taiwan Ocean University, Keelung 20224, Taiwan\\
	\email{lincc@mail.ntou.edu.tw}}
\maketitle              
\begin{abstract}
	The domination problem and its variants represent a classical domain within algorithmic graph theory. Among these variants, the paired-domination problem holds particular prominence due to its real-world implications in security and surveillance domains. Given an input graph $G$, the paired-domination problem involves identifying a minimum dominating set $D$ that induces a subgraph of $G$ with a perfect matching. Lin et al.~[\emph{Paired-domination problem on distance-hereditary graphs}, Algorithmica, 2020] previously presented a solution to this problem with a time complexity of $O(n^2)$.
	This paper significantly enhances their findings by introducing an $O(n+m)$-time algorithm. Furthermore, the time complexity of this algorithm can be reduced to $O(n)$ when provided with a decomposition tree for the graph $G$.
	
	\keywords{linear-time \and paired-domination \and distance-hereditary graphs~\and perfect matching \and decomposition tree \and dynamic programming.}
\end{abstract}

\baselineskip 14pt

\section{Introduction}\label{sec:introduction}
The investigation of the domination problem and its variants is a classical area in algorithmic graph theory, having undergone thorough investigation over the course of several decades. In a given graph $G=(V,E)$, a subset of vertices $D\subseteq V$ is regarded as a dominating set of $G$ if each vertex $v\in V$ either belongs to $D$ or is adjacent to a vertex within $D$. This fundamental problem, alongside its diverse adaptations, holds practical significance across various real-world domains, including electric power networks~\cite{Bjorkman20,Guo08,Haynes02}, network routing~\cite{Tian09,Wu01}, and resource allocation~\cite{Grinstead93,Shen13}.

One particularly famous variant from the domination problem emerges in the form of paired-domination, an innovation initially proposed by Haynes and Slater~\cite{HS98}. Their impetus for its inception stemmed from its tangible applications within security and surveillance contexts. In a museum protection program, the traditional domination problem necessitates the presence of a guard within each region or the encompassment of a region within the protective sphere of influence of a guard. However, the safeguarding of the guards themselves also demands consideration. In pursuit of heightened security, the guards have to back each other up, aligning precisely with the requisites of the paired-domination problem.

A vertex set $D$ is designated as a paired-dominating set if $D$ is a dominating set with subgraph $G[D]$ contains a perfect matching. The central aim of the paired-domination problem revolves around the identification of such a paired-dominating set characterized by the smallest cardinality. Given its categorization as NP-complete in general graphs~\cite{HS98}, this problem has captured the extensive interest of researchers across a spectrum of theoretical computer science domains, spanning approximation algorithms~\cite{Chang12,Chen09,Henning20}, hardness outcomes~\cite{Alvarado15,Alvarado16,Chen09}, attributes of the paired-domination number~\cite{Alvarado15,Henning10,Lu19}, and the formulation of polynomial-time algorithms tailored for pivotal graph classes~\cite{Chen2009,Kang04,Lappas13,Lin15,Qiao03}. In particular, Lin et al.~\cite{Lin15} devised a linear-time algorithm for addressing paired-domination within circular-arc graphs. Alvarado~\cite{Alvarado15} characterized classes of graphs exhibiting equivalent domination, total domination, and paired-domination numbers. Furthermore, Chen et al.~\cite{Chen09} introduced an approximation algorithm achieving a ratio of $\ln (2 \Delta) + 1$.

Broadening our scope of inquiry, we immerse ourselves in the realm of \emph{distance-hereditary graphs}. A graph is distance-hereditary if, within every connected induced subgraph, each pair of vertices maintains identical distances as in the original graph. This category finds its place as a subset of circle graphs and includes the domain of cographs. Over the past two decades, the attributes and optimization predicaments inherent to distance-hereditary graphs have been subjected to thorough analysis. For distance-hereditary graphs, including classical domination, independent domination, connected domination, and total domination, specialized $O(n+m)$-time algorithms have been devised~\cite{Chang1997,Chang02,Hsieh02,Yeh98}. Nevertheless, the paired-domination problem on distance-hereditary graphs is still bound by a time complexity of $O(n^2)$~\cite{lin2020}, representing a substantive focal point warranting further exploration. Therefore, the results presented in this paper provide extensive insights that enhance our understanding of the paired-domination challenge within the realm of distance-hereditary graphs.

The subsequent sections of this paper are structured as follows. Section \ref{section:algo-dynamic} introduces the notations, outlines the algorithm, and analyzes its time complexity. Section~\ref{section:algo} establishes the correctness of the algorithm and elucidates its detailed implementation. Ultimately, in Section~\ref{sec:conclusion}, we draw conclusions and propose potential avenues for future research.

\section{Algorithm for Distance-Hereditary Graphs}
\label{section:algo-dynamic}
This section presents a dynamic programming-based algorithm for efficiently determining a minimum paired-dominating set of a given distance-hereditary graph $G$ in $O(n+m)$ time. To start, we introduce essential notations. Suppose $G$ is an undirected connected graph, where $V(G)$ encompasses $n$ vertices and $E(G)$ consists of $m$ edges. For $v\in V(G)$, its \emph{open neighborhood} is denoted as $N(v)=\{u \mid (u,v) \in E(G)\}$, and the \emph{closed neighborhood} is defined as $N[v]=N(v)\cup\{v\}$. 
Moreover, a \emph{matching} $M \subseteq E(G)$ in graph $G$ is a set of edges where each vertex $v\in V(G)$ is incident to at most one edge in $M$. A \emph{perfect matching} is a matching in which every vertex appears exactly once. The subgraph of $G$ induced by a vertex set $S \subseteq V(G)$ is denoted by $G[S]$ and consists of the vertices in $S$ along with the edges in $\{(u,v)\in E(G) \mid u, v \in S\}$.
A subset of vertices $S\subseteq V$ is referred to as a \emph{dominating set} of graph $G$ if every vertex $v\in V$ is either an element of $S$ or adjacent to a vertex in $S$. A dominating set $S$ is regarded as a \emph{paired-dominating set} if the induced subgraph $G[S]$ contains a perfect matching. The size of a minimum paired-dominating set of $G$ is denoted as the \emph{paired-domination number} $\gamma_p(G)$. We denote by $D_p(G)$ the set of all paired-dominating sets of $G$ with cardinality equal to $\gamma_p(G)$.

Subsequently, we introduce the concept of the \emph{decomposition tree}, a data structure that encapsulates crucial attributes of distance-hereditary graphs. The decomposition tree guides our algorithm, which employs dynamic programming techniques to systematically traverse the vertices of $G$ and determine the minimum paired-dominating set. Finally, we provide an evaluation of the time complexity of our algorithm.
\subsection{Decomposition Trees}
\label{subsection:decomposition tree}
In their work \cite{Chang97}, Chang et al. showcased the representation of a distance-hereditary graph $G$ through an associated \emph{decomposition tree} denoted as $T$. This \emph{decomposition tree} $T$ adopts the structure of a perfect binary tree, where each rooted subtree corresponds to a distinct subgraph of $G$. The leaves of $T$ exhibit a one-to-one correspondence with the vertices of $G$, while each internal vertex $v$ of $T$ carries a label that elucidates the configuration of the subgraph of $G$ connected to $v$, concerning the subgraphs of its child nodes. The labels $\otimes$, $\odot$, and $\oplus$ symbolize specific operations: the \emph{true twin operation}, the \emph{false twin operation}, and the \emph{attachment operation}, respectively. These operations primarily concern the edge connections between two distinct sets of vertices known as \emph{twin sets}.
Initially, a graph featuring a sole vertex $v$ is regarded as a distance-hereditary graph, with its corresponding twin set designated as $TS(G)=\{v\}$. Subsequently, for two distance-hereditary graphs $G_l$ and $G_r$, accompanied by their respective twin sets $TS(G_l)$ and $TS(G_r)$, the specific details of each operation are illustrated as follows.

\vspace{3pt}

\begin{itemize}[itemsep=5pt]
	\item True twin operation (denoted as $G=G_l \otimes G_r$)\\
	\mbox{~}\hspace{3pt}$V(G)=V(G_l)\cup V(G_r)$.\\
	\mbox{~}\hspace{3pt}$E(G)=E(G_l)\cup E(G_r)\cup \{(u,v) \mid  u\in TS(G_l) ~\text{and}~  v\in TS(G_r)\}$.\\
	\mbox{~}\hspace{3pt}$TS(G)=TS(G_l)\cup TS(G_r)$.
	
	\item False twin operation (denoted as $G=G_l \odot G_r$)\\
	\mbox{~}\hspace{3pt}$V(G)=V(G_l)\cup V(G_r)$.\\
	\mbox{~}\hspace{3pt}$E(G)=E(G_l)\cup E(G_r)$.\\
	\mbox{~}\hspace{3pt}$TS(G)=TS(G_l)\cup TS(G_r)$.
	
	\item Attachment operation (denoted as $G=G_l \oplus G_r$)\\
	\mbox{~}\hspace{3pt}$V(G)=V(G_l)\cup V(G_r)$.\\
	\mbox{~}\hspace{3pt}$E(G)=E(G_l)\cup E(G_r)\cup \{(u,v) \mid  u\in TS(G_l)~\text{and}~ v\in TS(G_r)\}$.\\
	\mbox{~}\hspace{3pt}$TS(G)=TS(G_l)$.
\end{itemize}
\vspace{-15pt}

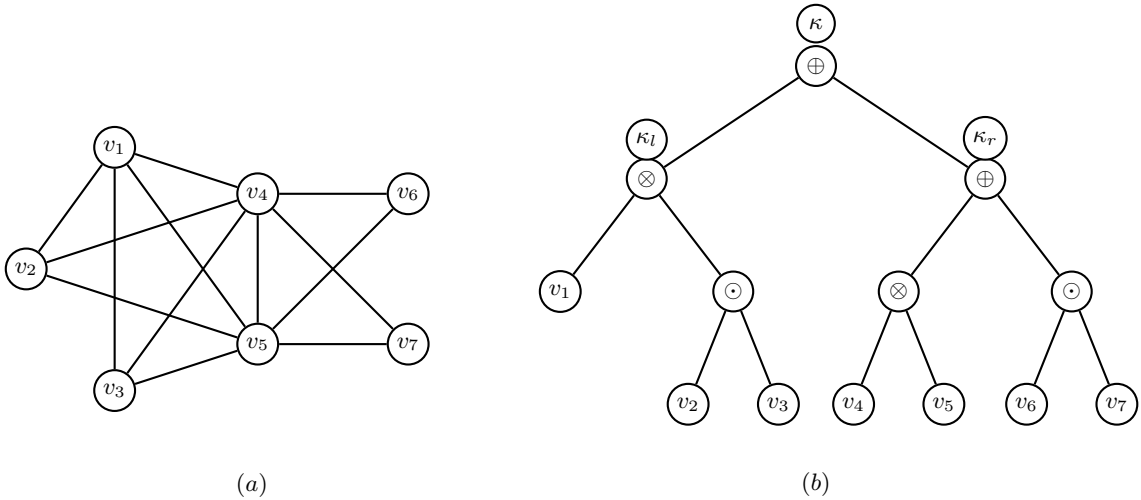
\begin{figure}[htb]
 	\hspace*{14pt}
	\hspace*{1em}{
\begin{tikzpicture}[level distance=1.3cm,
level 1/.style={sibling distance=4.5cm, level distance=1.5cm},
level 2/.style={sibling distance=2.3cm, level distance=1.5cm},
level 3/.style={sibling distance=1.2cm, level distance=1.5cm},
every node/.style={sibling distance=3cm, circle,   draw=black, fill=white, thick, minimum size=5mm, inner sep=2pt},
edge from parent/.style={sibling distance=3cm, black,thick,draw}]

\node[circle, draw=black, fill=white, thick, minimum size=5mm, inner sep=2pt] (1) at (0.588*2,0.809*2) {$v_1$};
\node[circle, draw=black, fill=white, thick, minimum size=5mm, inner sep=2pt] (2) at (0,0) {$v_2$};
\node[circle, draw=black, fill=white, thick, minimum size=5mm, inner sep=2pt] (3) at (0.588*2,-0.809*2) {$v_3$};
\node[circle, draw=black, fill=white, thick, minimum size=5mm, inner sep=2pt] (4) at (3.078,1) {$v_4$};
\node[circle, draw=black, fill=white, thick, minimum size=5mm, inner sep=2pt] (5) at (3.078,-1) {$v_5$};
\node[circle, draw=black, fill=white, thick, minimum size=5mm, inner sep=2pt] (6) at (5.078,1) {$v_6$};
\node[circle, draw=black, fill=white, thick, minimum size=5mm, inner sep=2pt] (7) at (5.078,-1) {$v_7$};
\draw[thick] (1) -- (2);
\draw[thick] (1) -- (3);
\draw[thick] (4) -- (6);
\draw[thick] (4) -- (7);
\draw[thick] (5) -- (6);
\draw[thick] (5) -- (7);
\draw[thick] (4) -- (1);
\draw[thick] (4) -- (2);
\draw[thick] (4) -- (3);
\draw[thick] (4) -- (5);
\draw[thick] (5) -- (1);
\draw[thick] (5) -- (2);
\draw[thick] (5) -- (3);

\node at (10.5,2.7) {$\oplus$}
child {node {$\otimes$}
    child {node {$v_1$}
    }
    child {node {$\odot$}
        child {node {$v_2$}}
        child {node {$v_3$}}
    }
}
child {node {$\oplus$}
    child {node {$\otimes$}
        child {node {$v_4$}}
        child {node {$v_5$}}
    }
    child {node {$\odot$}
        child {node {$v_6$}}
        child {node {$v_7$}}
    }
};

\node[below, draw=white, inner sep=-0pt] at (3,-2.6) {$(a)$};
\node[below, draw=white, inner sep=-0pt] at (10.5,-2.6) {$(b)$};
\draw (10.5,3) node [above, opacity=0, text opacity=1] {$\kappa$};
\draw (8.25,1.44) node [above, opacity=0, text opacity=1]
{$\kappa_l$}; \draw (12.75,1.44) node [above, opacity=0, text
opacity=1] {$\kappa_r$};
\end{tikzpicture}}
	\vspace{-12pt}
	\caption{A distance-hereditary graph $G$ with its corresponding decomposition tree $T$ rooted at $\kappa$.} 
	\label{figure:decomposition-tree} 
	\vspace{-10pt}
\end{figure}

Referring to Fig.\ref{figure:decomposition-tree}, the associated decomposition tree $T$ rooted at $\kappa$ illustrates the distance-hereditary graph $G$ as depicted in Fig.\ref{figure:decomposition-tree}($a$). Notably, the subgraph corresponding to the subtree rooted at $\kappa_l$ corresponds to $G[\{v_1,v_2,v_3\}]$ with a twin set of $\{v_1,v_2,v_3\}$. Similarly, the subgraph corresponding to the subtree rooted at $\kappa_r$ is $G[\{v_4,v_5,v_6,v_7\}]$ with a twin set of $\{v_4,v_5\}$. Given that the root $\kappa$ is labeled with $\otimes$, in accordance with the definition of the true twin operation, the subgraph corresponding to $\kappa$ encompasses the entire graph $G$ with a twin set of $\{v_1,v_2,v_3\}$. It was established by Chang et al. that a graph $G$ is distance-hereditary if and only if it possesses a decomposition tree. Furthermore, they introduced an algorithm for constructing $T$ in linear time.

\vspace{10pt}

\begin{Lemma}[\cite{Chang97,Hsieh02}] \label{lemma:decomposition-tree}
	Constructing a decomposition tree $T$ for a distance-hereditary graph $G$ can be accomplished in $O(n+m)$ time.
\end{Lemma}

\subsection{The Algorithm}
\label{subsection:algorithm}
For a given distance-hereditary graph $G$, our algorithm commences by establishing a decomposition tree $T$ specific to $G$. Following this, the algorithm employs a dynamic programming paradigm to systematically handle the vertices within $T$ through a bottom-up approach, resulting in a time complexity of $O(n)$. Before we go on to describe the algorithm in detail, several terminologies warrant elucidation.
Let $H$ be an induced subgraph of $G$. For $0 \le k \le |TS(H)|$, we define

\begin{enumerate}[label=$D_\arabic*(G):$, leftmargin=1.9cm, rightmargin=0.8cm, itemsep = 0pt]
	
	\item [$D_k(H)=$] \hspace{-5pt}
	$\{D \mid D$ is a minimum vertex subset over all $S\subseteq V(H)$ such that $V(H)-TS(H)\subseteq N_H[S]$, and $G[S-X]$ contains a perfect matching $M$ for some vertex subset $X\subseteq S \cap TS(H)$  with  $|X|=k\}$. 

\end{enumerate}

\noindent  The utility of the concept denoted as $D_k(H)$ becomes evident in its ability to infer $D_p(H)$, a parameter that frequently presents challenges when attempting direct determination. Let $\gamma_k(H)$ represent the cardinality of the set $D$ within the scope of $D_k(H)$. The algorithm we employ effectively calculates both $\gamma_k(H)$ and $\gamma_p(H)$ for each other in a methodical, hierarchical manner within a deconstructive tree structure.
In the present context, the vertices encompassed by set $X$ are designated as \emph{unpaired} in relation to the pair $(D, M)$. In the context of a given vertex $v \in T$, the notation $\hat{V}(v)$ indicates the assemblage of terminal nodes within the substructure rooted at vertex $v$. For convenience, we introduce the following definitions: $\hat{D}_k(v)=D_k(G[\hat{V}(v)])$, $\hat{D}_p(v) = D_p(G[\hat{V}(v)])$, $\hat{\gamma}_k(v)=\gamma_k(G[\hat{V}(v)])$, $\hat{\gamma}_p(v) = \gamma_p(G[\hat{V}(v)])$, $\hat{E} (v) = E(G[\hat{V}(v)])$, $\hat{TS}(v) = TS(G[\hat{V}(v)])$, and $\hat{G}(v) = G[\hat{V}(v)]$. An illustrative instance is provided in Fig.~\ref{figure:decomposition-tree}. Given $\hat{TS}(\kappa) = \{v_1,v_2,v_3,v_4,v_5\}$, it can be confirmed that $\{v_5, v_6\} \in \hat{D}_0(\kappa)$, $\{v_4\} \in \hat{D}_1(\kappa)$, $\{v_1, v_4\} \in \hat{D}_2(\kappa)$, $\{v_1, v_2, v_4\} \in \hat{D}_3(\kappa)$, $\{v_1, v_2, v_3, v_4\} \in \hat{D}_4(\kappa)$, and $\{v_1, v_2, v_3, v_4, v_5\}\in \hat{D}_5(\kappa)$. Furthermore, the vertex set $\{v_3, v_4\}$ constitutes a minimal paired-dominating set of $G$, yielding $\gamma_p(G) = \hat{\gamma}_p(\kappa) = 2$.

In addition, we introduce a binary variable denoted as $\emptp(v)$, which takes on the value of $1$ exclusively in cases where $\hat D_p(v) \cap \hat D_0(v) = \emptyset$. With these notational components established, we proceed to outline the recursive definition of $\hat{\gamma}_p(v)$, undertaken in a two-stage process. Let us assume that vertex $v$ is an internal node within the structure $T$, with $v_l$ and $v_r$ representing its left and right children respectively. The initial recursive equation is 

\vspace{-0.5cm}
\begin{center}
	\begin{equation}\label{eq1}
		\hat\gamma_p(v)=
		\begin{cases}
			\hat\gamma_p(v_l)+\hat\gamma_p(v_r)    &\text{if $v$ is denoted by $\odot$},\\        
			\hat\gamma_{0}(v)+ 2*\emptp(v)   &\text{otherwise.}
		\end{cases}
	\end{equation}
\end{center}

\vspace{0.2cm}

\noindent Evidently, the formulation of the recursive equation is contingent upon the label attributed to vertex $v$, thereby molding the computational process for $\hat{\gamma}_p(v)$. In instances where the label assigned to $v$ assumes the value $\odot$, the focus pivots towards the subproblems entailing the determination of $\hat{\gamma}_p(v_l)$ and $\hat{\gamma}_p(v_r)$. Conversely, when $v$ is characterized by $\otimes$ or $\oplus$, the attention shifts towards resolving the intricacies inherent in $\hat{\gamma}_0(v)$ and $\emptp(v)$.

\smallskip
\begin{Lemma} \label{gamma_p(G)=gamma_p(G)+b_all}
	For every internal vertex $v$ within $T$, Equation~(\ref{eq1}) holds.
\end{Lemma}
\begin{proof}
	In scenarios where vertex $v$ is labeled as $\odot$, the equation $\hat{\gamma}_p(v) = \hat{\gamma}_p(v_l) + \hat{\gamma}_p(v_r)$ holds true due to the absence of any edge $(x, y) \in E(G)$ where $x \in \hat{V}(v_l)$ and $y \in \hat{V}(v_r)$. Consequently, we consider the instances where $v$ carries the labels $\otimes$ or $\oplus$. Given that $\hat{V}(v) - \hat{TS}(v) \subseteq \hat{V}(v)$, it follows that $\hat{\gamma}_p(v) \ge \hat{\gamma}_0(v)$. When confronted with the situation where $\hat D_p(v) \cap \hat D_0(v) \neq \emptyset$, it becomes evident that $\hat{\gamma}_p(v) = \hat{\gamma}_0(v)$, thus solidifying the equation's validity.
	Subsequently, the focus shifts to instances where $\hat D_p(v) \cap \hat D_0(v) = \emptyset$, thereby implying $\hat{\gamma}_p(v) > \hat{\gamma}_0(v)$, which further leads to $\hat{\gamma}_p(v) \ge \hat{\gamma}_0(v) + 2$. Remarkably, the $\otimes$ operation displays symmetry. Without loss of generality, let us assume that $D \in \hat{D}_0(v)$ such that $\hat{TS}(v_l) \not \subseteq N_{\hat G(v)}[D]$. Since $v$ bears the labels $\otimes$ or $\oplus$, it follows that $\hat{TS}(v_r) \cap D = \emptyset$. Let $x$ and $y$ denote the vertices in $\hat{TS}(v_l) - D$ and $\hat{TS}(v_r)$ respectively. It is evident that $D \cup \{x, y\}$ retains its status as a paired-dominating set of $G$, thereby yielding the conclusion $\hat{\gamma}_p(v) \le \hat{\gamma}_0(v) + 2$. With these observations, the lemma's proof attains its completion.
\end{proof}

\vspace{1pt}

As Equation~(\ref{eq1}) indicates, determining $\hat\gamma_p(v)$ involves finding $\hat\gamma_{0}(v)$ and $\emptp(v)$ when $v$ bears the labels $\otimes$ or $\oplus$. The boolean variable $\empts(v)$ is assigned true exclusively when $D\cap\hat{TS}(v) = \emptyset$ holds for all sets $D$ within $\hat{D}_0(v)$. In the forthcoming Subsection~\ref{subsection:emptp_empts}, assuming that $\emptp(v_i)$ and $\empts(v_i)$ are known for $i \in \{l,r\}$, we shall establish that the calculation of $\emptp(v)$ can be executed in $O(1)$ time.
However, the direct determination of $\hat{\gamma}_0(v)$ remains challenging due to the myriad combinations of $D_l \in \hat{D}_x(v_l)$ and $D_r \in \hat{D}_y(v_r)$ that give rise to $D \in \hat{D}_0(v)$, subject to $0\le x \le |\hat{TS}(v_l)|$ and $0\le y \le |\hat{TS}(v_r)|$. For instance, a dominating set $D \in \hat{D}_0(v)$ could potentially emerge from certain $D_l \in \hat{D}_k(v_l)$ and $D_r \in \hat{D}_k(v_r)$ configurations where $D = D_l \cup D_r$, while adhering to $0 \le k\le \min \{ |\hat{TS}(v_l)|,|\hat{TS}(v_r)|\}$. Exhaustively exploring all feasible combinations incurs considerable time costs. To effectively tackle this challenge, the second recursive equation presents a swift strategy for determining $\hat{\gamma}_k(v)$ for $0\le k \le |\hat{TS}(v)|$, encompassing $\hat{\gamma}_0(v)$.
Let $\hat{\text{min}}(v) = \min \{ \hat{\gamma}_k(v) \mid 0\le k\le |\hat{TS}(v)| \}$ and $\Psi(v) = \{ D \mid D \in \hat{D}_k(v) \text{ such that } |D| = \hat{\text{min}}(v) \}$. Moreover, introduce $\hat{\alpha}(v)$ to denote the minimum value and $\hat{\beta}(v)$ to represent the maximum value of $k$ such that $\hat{\gamma}_{\hat{\alpha}(v)}(v) = \hat{\text{min}}(v)$ and $\hat{\gamma}_{\hat{\beta}(v)}(v) = \hat{\text{min}}(v)$, respectively. These preliminaries pave the way for the second recursive equation:

\vspace{-0.49cm}
\begin{center}  
	\begin{equation}\label{eq2}
		\hat\gamma_k(v)=
		\begin{cases}
			\hat{min}(v) + \hat\alpha(v)-k         &\text{if $0 \le k \le \hat\alpha(v)$},\\
			\hat{min}(v) + k - \hat\beta(v)        &\text{if $\hat\beta(v) \le k \le |\hat{TS}(v)|$},\\
			\hat{min}(v)                           &\text{if $\hat\alpha(v) < k < \hat\beta(v)$ and $k-\hat\alpha(v)$ is even},\\
			\hat{min}(v) + 1                       &\text{otherwise.}
		\end{cases}
	\end{equation}
\end{center}

\vspace{0.2cm}
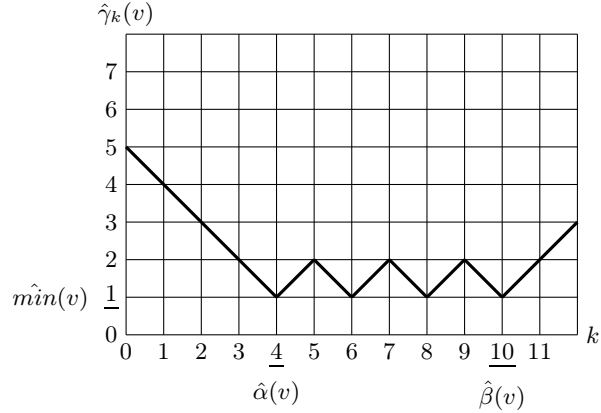
\begin{figure} 
	\begin{center}
		\hspace*{-5em}{
\begin{tikzpicture}
\foreach \x in {0, 0.5, ...,6}
\draw (\x,0) -- (\x,4);

\foreach \y in {0, 0.5, ...,4}
\draw (0,\y) -- (6,\y);

\draw[scale=1, domain=0:1, very thick] (0,2.5) -- (2,0.5);
\draw[scale=1, domain=0:1, very thick] (2,0.5) -- (2.5,1);
\draw[scale=1, domain=0:1, very thick] (2.5,1) -- (3,0.5);
\draw[scale=1, domain=0:1, very thick] (3,0.5) -- (3.5,1);
\draw[scale=1, domain=0:1, very thick] (3.5,1) -- (4,0.5);
\draw[scale=1, domain=0:1, very thick] (4,0.5) -- (4.5,1);
\draw[scale=1, domain=0:1, very thick] (4.5,1) -- (5,0.5);
\draw[scale=1, domain=0:1, very thick] (5,0.5) -- (6,1.5);

\foreach \x in {0, 1, 2,3,5,6,7,8,9, 11}
\draw (\x/2,0) node [below] {\x};
\draw (2,0) node [below] {\ul{4}};
\draw (5,0) node [below] {\ul{10}};

\foreach \y in {0, 2, 3, 4, 5, 6, 7}
\draw (0,\y/2) node [left] {\y};
\draw (0,0.5) node [left] {\ul{1}};

\draw (0,4) node [left,above] {$\hat \gamma_k(v)$};
\draw (6,0) node [right] {$k$};

\draw (-0.38,0.5) node [left] {$\hat{min}(v)$};
\draw (2,-0.5) node [below] {$\hat\alpha(v)$};
\draw (5,-0.5) node [below] {$\hat\beta(v)$};
\end{tikzpicture}}
		\vspace{-20pt}
		\caption{The relationship between the unpaired vertex numbers $k$ and their corresponding $\hat\gamma_k(v)$ of $\hat G(v)$.} \label{figure:function}
	\end{center}
	\vspace{-17pt}
\end{figure}

Before delving into the proof of correctness, we pause to observe certain properties inherent in the equation, as depicted in Fig.~\ref{figure:function}. It is evident that when $k \le \hat{\alpha}(v)$, the equation exhibits a slope of $-1$. Consequently, an incremental adjustment of $k$ by a unit results in a corresponding reduction of $\hat{\gamma}_k(v)$ by one unit. Conversely, when $\hat{\beta}(v) \le k$, the equation takes on a slope of $1$, maintaining symmetry. Furthermore, within the range $\hat{\alpha}(v) < k < \hat{\beta}(v)$, $\hat{\gamma}_k(v)$ adopts values of either $\hat{\text{min}}(v)$ or $\hat{\text{min}}(v) + 1$. The equation manifestly elucidates the interplay between $k$ and $\hat{\gamma}_k(v)$. This clarity underscores the notion that $\hat{\gamma}_k(v)$ can be determined within $O(1)$ time, given $\hat{\text{min}}(v)$, $\hat{\alpha}(v)$, and $\hat{\beta}(v)$ as inputs. This strategic division enables us to disentangle the intricate computation into smaller components.

\smallskip
\begin{Lemma} \label{lemma-ep2}
	For every internal vertex $v$ within $T$, Equation~(\ref{eq2}) holds.
\end{Lemma}
\begin{proof}
	We proceed to prove the lemma through induction based on the height of vertex $v$ within the structure $T$. If the height of $v$ is $0$, then $v$ must be a leaf, implying $\hat\gamma_0(v) = 0$ and $\hat\gamma_1(v) = 1$. Consequently, we have $\hat{\text{min}}(v) = 0$, $\hat\alpha(v) = 0$, and $\hat\beta(v) = 0$, thereby confirming the lemma's validity in this case.
	Next, let's assume that $v$ is an internal vertex with left child $v_l$ and right child $v_r$. By applying the induction hypothesis, Equation~(\ref{eq2}) holds for both $v_l$ and $v_r$. Under these premises, the assertions provided by Lemmas~\ref{lemma:eq2_holds_for_true}, \ref{lemma:eq2_holds_for_false}, and~\ref{lemma:eq2_holds_for_attachment} presented in Section~\ref{section:algo} ascertain the lemma's validity for instances where $v$ bears the labels corresponding to the true twin operation $\otimes$, false twin operation $\odot$, and attachment operation $\oplus$, respectively. Consequently, the lemma stands affirmed.
\end{proof}


Equipped with Equations~(\ref{eq1}) and~(\ref{eq2}), we now proceed to provide our algorithm, which efficiently computes $\hat\gamma_{p}(v)$, $\hat{\text{min}}(v)$, $\hat\alpha(v)$, $\hat\beta(v)$, $\empts(v)$, and $\emptp(v)$ in a bottom-up manner. The algorithm commences by constructing a decomposition tree $T$ for graph $G$, designating $\kappa$ as the root, and labeling all vertices as unvisited.
For each leaf $\ell \in T$, we set $\hat\gamma_{p}(\ell) = \infty$, $\hat{\text{min}}(\ell) = 0$, $\hat\alpha(\ell) = 0$, $\hat\beta(\ell) = 0$, $\empts(\ell) = 1$, and $\emptp(\ell) = 1$, subsequently marking $\ell$ as visited. Subsequently, the algorithm advances through iterations, progressively processing internal vertices within $T$. In each iteration, we select a vertex $v$ from $T$ whose child vertices have both been marked as visited. This criterion signifies that the values $\hat\gamma_{p}(v_i)$, $\hat{\text{min}}(v_i)$, $\hat\alpha(v_i)$, $\hat\beta(v_i)$, $\empts(v_i)$, and $\emptp(v_i)$ for $i \in \{l,r\}$ have been established.
With these values in hand, we proceed to showcase $O(1)$-time procedures for computing $\hat{\text{min}}(v)$, $\hat\alpha(v)$, and $\hat\beta(v)$, which are detailed in Subsection~\ref{subsection:true_twin}, Subsection~\ref{subsection:false_twin}, and Subsection~\ref{subsection:attachment}, respectively. Each of these subsections corresponds to the true twin $\otimes$, false twin $\odot$, and attachment $\oplus$ operations. Additionally, the methodology for determining $\empts(v)$ and $\emptp(v)$ is expounded upon in Subsection~\ref{subsection:emptp_empts}.
Following the computation of $\hat\gamma_{0}(v)$ and $\emptp(v)$, we proceed to calculate $\hat\gamma_{p}(v)$ using Equation~(\ref{eq1}). Once all vertices have been visited and marked accordingly, the algorithm concludes by providing the paired-dominating number $\hat\gamma_{p}(\kappa)$ for graph $G$. A comprehensive depiction of the algorithm is presented in Algorithm~\ref{algo:main}.

\begin{algorithm}[htb]
	\caption{Determining the cardinality of minimum paired-dominating set on distance-hereditary graphs}\label{algo:main}
	\begin{algorithmic} [1]
		\baselineskip 14pt
		\REQUIRE a distance-hereditary graph $G$.
		\ENSURE  the paired-dominating number $\gamma_{p}(G)$.
		
		\STATE build a decomposition tree $T$ with $\kappa$ as its root for $G$;
		\STATE set all vertices as unvisited; 
		\FOR{every leaf $\ell \in T$}
		\STATE let $\hat\gamma_{p}(\ell) \leftarrow \infty$, $\hat{min}(\ell) \leftarrow 0$, $\hat\alpha(\ell) \leftarrow 0$, $\hat\beta(\ell) \leftarrow 0$, $\empts(\ell) \leftarrow 1$, and $\emptp(\ell) \leftarrow 1$;
		\STATE set $\ell$ as visited;
		\ENDFOR
		
		\REPEAT
		\STATE select a vertex $v$ of $T$ with both of its children $v_l$ and $v_r$ have been marked as visited;
		\STATE compute $\hat\gamma_{p}(v)$, $\hat{min}(v)$, $\hat\alpha(v)$, $\hat\beta(v)$, $\empts(v)$, and $\emptp(v)$ using the previously determined values of $v_l$ and $v_r$ in earlier iterations;
		\STATE set $v$ as visited;
		\UNTIL{the root $\kappa$ of $T$ is marked as visited}
		
		\RETURN $\hat\gamma_{p}(\kappa)$.
	\end{algorithmic}
\end{algorithm}

\vspace{6pt}
\baselineskip 14pt
With reference to Equations~(\ref{eq1}) and~(\ref{eq2}), the correctness of Algorithm~\ref{algo:main} can be established through the verification of the procedures that correctly compute the values $\hat{\text{min}}(v)$, $\hat\alpha(v)$, $\hat\beta(v)$, $\empts(v)$, and $\emptp(v)$. This verification process will be carried out by Lemmas~\ref{lemma:determine_values_for_true}, \ref{lemma:determine_values_for_false}, \ref{lemma:determine_values_for_attachment}, \ref{lemma:determine_emptp_empts_for_true}, \ref{lemma:determine_emptp_empts_for_false}, and \ref{lemma:determine_emptp_empts_for_attach}.
Subsequently, the algorithm's time complexity is analyzed below. Given that each internal vertex can be processed in $O(1)$ time and that $T$ comprises precisely $n-1$ internal vertices, the runtime of the loop at Steps $(7)$--$(11)$ is bounded by $O(n)$. As per Lemma~\ref{lemma:decomposition-tree}, a decomposition tree $T$ for graph $G$ can be constructed in $O(n+m)$ time. Meanwhile, since the other steps can also be executed within $O(n)$ time, the worst-case runtime of Algorithm~\ref{algo:main} is $O(n + m)$. Furthermore, during the algorithm's execution, it becomes possible to construct a minimum paired-dominating set by recording the choices that lead to the optimal value $\hat{\gamma}_{p}(\kappa)$. Thus, the principal result of this study is established.

\begin{theorem}\label{lemma:main-result}
	Given a distance-hereditary graph $G$, a minimum paired-dominating set of $G$ can be determined in $O(n + m)$ time.
\end{theorem}


\section{Correctness and Implementation Details}
\label{section:algo}
Certainly, in this section, we focus on an internal vertex $v$ of decomposition tree $T$ for graph $G$. It is assumed that $v$ has a left child $v_l$ and a right child $v_r$. Additionally, we further assume that Equation~(\ref{eq2}) holds true for both $v_l$ and $v_r$. Furthermore, the values $\hat{\text{min}}(v_i)$, $\hat\alpha(v_i)$, $\hat\beta(v_i)$, $\empts(v_i)$, and $\emptp(v_i)$ have been successfully established for $i \in \{l,r\}$. This section aim to establish the validity of Equation~(\ref{eq2}) for vertex $v$, as well as to showcase the capability to calculate $\hat{\text{min}}(v)$, $\hat\alpha(v)$, and $\hat\beta(v)$ within a $O(1)$ timeframe.

To achieve this, the discussion is structured into three subsections, each aligned with the label attributed to vertex $v$: true twin $\otimes$, false twin $\odot$, and attachment $\oplus$ operations. These subsections are elaborated upon in Subsections~\ref{subsection:true_twin}, \ref{subsection:false_twin}, and \ref{subsection:attachment}, respectively.

\vspace{5pt}
\subsection{Key properties of $\hat\gamma_k(v)$} 
\label{subsection:properties} 
Within this subsection, we demonstrate the fact that the disparity between $\hat\gamma_k(v)$ and $\hat\gamma_{k+1}(v)$ amounts precisely to $1$, as visually depicted in Fig~\ref{figure:function}. This particular property not only holds considerable significance for forthcoming discussions but also offers a novel approach for proving Equation~(\ref{eq2}).

\begin{Lemma} \label{lemma:k=k+-1}
	$|\hat\gamma_k(v) - \hat\gamma_{k+1}(v)| = 1$ for $0\leq k < |\hat{TS}(v)|$. 
\end{Lemma}
\begin{proof} 
	The objective is to prove the inequality $\hat \gamma_{k+1}(v) - 1 \le \hat \gamma_k(v) \le \hat \gamma_{k+1}(v) + 1$ since the premise that the sum of $\hat\gamma_k(v)$ and $\hat\gamma_{k+1}(v)$ is an odd value. This assertion naturally leads to the conclusion that $\hat\gamma_k(v) \neq \hat\gamma_{k+1}(v)$.
	To begin, let us consider the case $\hat{TS}(v) \not \subseteq D$, where $D\in\hat D_k(v)$ and $X\in\hat{TS}(v)\cap D$ such that $|X| = k$. In this context, a vertex $u$ exists in $\hat{TS}(v) - D $. It can be verified that $D\cup \{u\}$ is a dominating set of $\hat{V}(v)-\hat{TS}(v)$, while $X \cup \{u\}$ remains unpaired with respect to $(D,M)$. This follows $\hat \gamma_{k+1}(v) \le \hat \gamma_k(v) + 1$.
	Moving on to the scenario where $\hat{TS}(v) \subseteq D$, we consider the possibility of an edge $(x,y) \in M$ with $x \in \hat{TS}(v)$. Firstly, let us examine the situation where $N_{\hat G(v)}(y) \not \subseteq D$. In this subcase, let $z$ be a vertex in $N_{\hat G(v)}(y) - D $. Here, the set $D \cup \{z\}$ functions as a dominating set for $\hat{V}(v)-\hat{TS}(v)$, preserving the unpaired nature of the vertices in $X \cup \{x\}$ relative to $(D\cup\{z\}, M \cup \{(y,z)\} - \{(x,y)\} )$.
	Alternatively, when $N_{\hat G(v)}(y) \subseteq D$, it is evident that $D - \{y\}$ constitutes a dominating set for $\hat{V}(v)-\hat{TS}(v)$, while the vertices in $X \cup \{x\}$ remain unpaired with respect to $(D-\{y\}, M - \{(x,y)\})$.
	In both subcases, the relationship $\hat \gamma_{k+1}(v) \le \hat \gamma_k(v) + 1$ is upheld. By employing similar lines of reasoning, the converse inequality $\hat \gamma_k(v) \le \hat \gamma_{k+1}(v)+1$ can also be established. However, for the sake of brevity, we shall omit the detailed elaboration of this process.
\end{proof}

\smallskip

Building upon the results of Lemma~\ref{lemma:k=k+-1}, we are now equipped to introduce an innovative approach for substantiating the validity of Equation~(\ref{eq2}).

\smallskip
\begin{Corollary} \label{corollary:proof_strategy}
	Equation~(\ref{eq2}) holds if the following three statements are true:
	\begin{enumerate}  \itemsep 0.5pt  
		\item $\hat\gamma_0(v) = \hat{min}(v) + \hat\alpha(v)$; 
		\item $\hat\gamma_{|\hat{TS}(v)|}(v)=\hat{min}(v)+ |\hat{TS}(v)|-\hat\beta(v)$; and
		\item $\hat\gamma_{\hat\alpha(v)+2i}(v) = \hat{min}(v)$ for $0 \le i \le (\hat\beta(v)-\hat\alpha(v))/2$.
	\end{enumerate}  
\end{Corollary}
\begin{proof}
	Leveraging the insights from Lemma~\ref{lemma:k=k+-1}, the process of establishing Equation~(\ref{eq2}) can be dissected into three distinct cases contingent on the index value $k$. When $0 \le k \le \hat\alpha(v)$, it is evident that $\hat\gamma_k(v) = \hat{min}(v) + \hat\alpha(v) - k$ if and only if $\hat\gamma_0(v) = \hat{min}(v) + \hat\alpha(v)$. Similarly, for $\hat\beta(v) \le k \le |\hat{TS}(v)|$, it can be deduced that $\hat\gamma_k(v) = \hat{min}(v) + k - \hat\beta(v)$ if and only if $\hat\gamma_{|\hat{TS}(v)|}(v) = \hat{min}(v) + |\hat{TS}(v)| - \hat\beta(v)$.
	Lastly, for the remaining cases of Equation~(\ref{eq2}), the statement holds true if and only if $\hat\gamma_{\hat\alpha(v) + 2i}(v) = \hat{min}(v)$ for $0 \le i \le (\hat\beta(v) - \hat\alpha(v))/2$.
\end{proof}


\medskip

\subsection{True Twin Operation $(G=G_l \otimes G_r)$} 
\label{subsection:true_twin}
In this subsection, we assume that an internal vertex $v$ of $T$ is labeled by true twin operation $\otimes$ with left child $v_l$ and right child $v_r$. A procedure, which is developed for determining $\hat{min}(v)$, $\hat\alpha(v)$, and $\hat\beta(v)$, is first presented. Then, we show that Equation~(\ref{eq2}) holds for $v$ by applying the strategy mentioned in Corollary~\ref{corollary:proof_strategy}.

\subsubsection{Determine $\hat{min}(v)$, $\hat\alpha(v)$, and $\hat\beta(v)$ for true twin operation} \label{subsection:determine_domination-number_for_true_twin}
\ 
\vspace{9pt}
\newline
Below, we first propose a procedure for finding $\hat{min}(v)$, $\hat\alpha(v)$, and $\hat\beta(v)$ and then present lemmas for proving the correctness. Each step of the procedure is simple and easy to implement. 


\floatname{algorithm}{Procedure}
\begin{algorithm}[htb]
	\caption{Determine the values $\hat{min}(v)$, $\hat\alpha(v)$, and $\hat\beta(v)$ for true twin operation $\otimes$}
	\label{algo:determine_values_for_true}
	\begin{algorithmic} [1]
		\baselineskip 14pt
		\REQUIRE $\hat{min}(v_i)$, $\hat\alpha(v_i)$, and $\hat\beta(v_i)$ with $i \in \setof{l,r}$.
		
		\ENSURE $\hat{min}(v)$, $\hat\alpha(v)$, and $\hat\beta(v)$.
		
		\STATE let $\hat{min}(v) \leftarrow \hat{min}(v_l) + \hat{min}(v_r)$;
		
		\STATE let $\hat\alpha(v) \leftarrow \max \setof{\hat\alpha(v_l)-\hat\beta(v_r), \hat\alpha(v_r)-\hat\beta(v_l), |\hat\alpha(v_l)-\hat\alpha(v_r)| \hspace{-5pt } \mod 2}$;
		
		\STATE let $\hat\beta(v) \leftarrow \hat \beta(v_l) + \hat \beta(v_r)$;
		
		\RETURN $\hat{min}(v)$, $\hat\alpha(v)$, and $\hat\beta(v)$.
	\end{algorithmic}
\end{algorithm} 

\baselineskip 14pt

Suppose that $D\in \hat D_k(v)$ and $X \subseteq D \cap \hat{TS}(v)$ is a vertex set such that $G[D - X]$ has a perfect matching $M$ with $|X| = k$. The number of unpaired vertices in $\hat{TS}(v_l)$ and $\hat{TS}(v_r)$ are $k_l = |X \cap \hat{TS}(v_l)|$ and $k_r = |X \cap \hat{TS}(v_r)|$, respectively. Suppose further that $M$ contains exactly $h$ edges $(x_1, y_1), (x_2, y_2),\ldots, (x_h, y_h)$ such that $x_i \in \hat{TS}(v_l)$ and $y_i \in \hat{TS}(v_r)$ for $1 \le i \le h$. Meanwhile, we set $M_l = M \cap \hat{E}(v_l)$, $M_r = M \cap \hat{E}(v_r)$, $X_l = (X \cap \hat{TS}(v_l)) \cup \setof{x_1,x_2,\ldots,x_h}$, and $X_r = (X \cap \hat{TS}(v_r)) \cup \setof{y_1,y_2,\ldots,y_h}$.

\smallskip

\begin{Lemma} \label{lemma:r_and_l_are_minimized_for_true}
	Suppose that $D\in \hat D_k(v)$, $D_l = D \cap \hat{V}(v_l)$ and $D_r = D \cap \hat{V}(v_r)$. Then, $D_l \in \hat D_{h + k_l}(v_l)$ and $D_r \in \hat D_{h + k_r}(v_r)$. Furthermore, $X_l$ and $X_r$ are two sets of unpaired vertices with respect to $(D_l, M_l)$ and $(D_r, M_r)$, respectively.
\end{Lemma}
\begin{proof}
	Clearly, $G[D_l - X_l]$ contains a perfect matching $M_l$ with $|X_l| = k_l + h$. Since there exists no edge $(x, y) \in \hat E(v)$ such that $x \in \hat{V}(v_l) - \hat{TS}(v_l)$ and $y \in D_r$, $D_l$ is a dominating set of  $\hat{V}(v_l)-\hat{TS}(v_l)$. Therefore, to prove $D_l \in \hat D_{h + k_l}(v_l)$, it suffices to show that the number of vertices in $D_l$ is minimized. Suppose for the purpose of contradiction that $S_l \in \hat D_{h + k_l}(v_l)$ and $|S_l| < |D_l|$. We further suppose that $G[S_l-X'_l]$ contains a perfect matching such that $X'_l = \setof{z_1,z_2,\ldots,z_{h+k_l}}$. Then, one can verify that $S_l \cup D_r$ is also a dominating set of  $\hat{V}(v)-\hat{TS}(v)$. Let $X' = \setof{z_1,z_2,\ldots,z_{k_l}} \cup (X_r - \{y_1,\ldots,y_{h}\})$. Since $(x,y) \in \hat E(v)$ for $ x \in \hat{TS}(v_l)$ and $y \in \hat{TS}(v_r)$, one can see that $G[\{S_l \cup D_r\}-X']$ contains a perfect matching with $|X'|= k_l + k_r = k$. This contradicts the fact $D\in \hat D_k(v)$ as $|S_l| + |D_r| < |D|$. Symmetrically, it can be shown that $D_r \in \hat D_{h + k_r}(v_r)$ and $G[D_r - X_r]$ contains a perfect matching $M_r$ with $|X_r| = k_r + h$.
\end{proof}

\begin{Lemma} \label{lemma:determine_min_for_true}
	$\hat{min}(v)=\hat{min}(v_l)+\hat{min}(v_r)$.
\end{Lemma}
\begin{proof}
	We first show that $\hat{min}(v) \ge \hat{min}(v_l)+\hat{min}(v_r)$. Recall that $\Psi(v) = \{D \mid D \in \hat{D}_k(v)$ for some $k$ and  $|D| = \hat{min}(v)\}$. Suppose $D\in \Psi(v)$, $D_l = D \cap \hat{V}(v_l)$, and $D_r = D \cap \hat{V}(v_r)$. Then, by Lemma~\ref{lemma:r_and_l_are_minimized_for_true}, we have $D_l \in \hat D_s(v_l)$ and $D_r \in \hat D_t(v_r)$ for some $0 \le s \le |\hat{TS}(v_l)|$ and $0 \le t \le |\hat{TS}(v_r)|$. Hence, $|D_l|= \hat\gamma_{s}(v_l) \ge \hat{min}(v_l)$ and $|D_r|= \hat\gamma_{t}(v_r) \ge \hat{min}(v_r)$. It follows that $\hat{min}(v) = |D| = |D_l|+ |D_r| \ge \hat{min}(v_l)+\hat{min}(v_r)$. Next, we show that $\hat{min}(v) \le \hat{min}(v_l)+\hat{min}(v_r)$. Suppose that $D_l \in \Psi(v_l)$ and $D_r \in \Psi(v_r)$ are two vertex sets such that $G[D_l - X_l]$ and $G[D_r - X_r]$ have perfect matchings $M_l$ and $M_r$, respectively. Then, $D = D_l \cup D_r$ is dominating set of $\hat{V}(v) - \hat{TS}(v)$ such that $G[D - (X_l \cup X_r)]$ has a perfect matchings $M_l \cup M_r$. Hence, we have $\hat{min}(v) \le |D| = \hat{min}(v_l) + \hat{min}(v_r)$.
\end{proof}

\smallskip
Before proceeding to prove the correctness of $\hat\alpha(v)$ and $\hat\beta(v)$, we require the following auxiliary lemma. 

\begin{Lemma} \label{lemma:true min merge}
	Suppose that $D\in \hat{V}(v)$, $D_l = D \cap \hat{V}(v_l)$ and $D_r = D \cap \hat{V}(v_r)$. Then, 
	we have $D\in \Psi(v)$ if and only if $D_l \in \Psi(v_l)$ and $D_r \in \Psi(v_r)$.
\end{Lemma}
\begin{proof} 
	First we show the sufficient condition. Suppose that $D\in \Psi(v)$. Then, by Lemma~\ref{lemma:r_and_l_are_minimized_for_true}, we have $D_l \in \hat D_s(v_l)$ and $D_r \in \hat D_t(v_r)$ for some $0 \le s \le |\hat{TS}(v_l)|$ and $0 \le t \le |\hat{TS}(v_r)|$. Hence, it remains to show $\hat D_s(v_l) \subseteq \Psi(v_l)$ and $\hat D_t(v_r) \subseteq \Psi(v_r)$. By Lemma~\ref{lemma:determine_min_for_true}, we have $|D_l| + |D_r| = |D| = \hat{min}(v_l) + \hat{min}(v_r)$. In addition, since $|D_l|\ge \hat{min}(v_l)$ and $|D_r|\ge \hat{min}(v_r)$, we have $|D_l| = \hat{min}(v_l)$ and $|D_r| = \hat{min}(v_r)$. This implies that $\hat D_s(v_l) \subseteq \Psi(v_l)$ and $\hat D_t(v_r) \subseteq \Psi(v_r)$.
	
	Let us go to the necessity. Suppose that $D_l \in \Psi(v_l) \cap \hat D_s(v_l)$ and $D_r \in \Psi(v_r) \cap \hat D_t(v_r)$ are two vertex sets such that $G[D_l - X_l]$ and $G[D_r - X_r]$ have perfect matchings $M_l$ and $M_r$ with $|X_l| = s$ and $|X_r| = t$, respectively. Then, $D = D_l \cup D_r$ is dominating set of $\hat{V}(v) - \hat{TS}(v)$ such that $G[D - (X_l \cup X_r)]$ has a perfect matching  $M_l \cup M_r$. Further, because of $|D| = |D_l| + |D_r| = \hat{min}(v_l) + \hat{min}(v_r) = \hat{min}(v)$, we have $D\in \Psi(v)$.
\end{proof}

\begin{Lemma} \label{lemma:determine_alpha_for_true}
	$
	\hat\alpha(v) =
	\begin{cases}
		\hat \alpha(v_l)- \hat \beta(v_r) & \text{if $\hat \alpha(v_l)> \hat \beta(v_r)$},\\
		\hat \alpha(v_r)- \hat \beta(v_l) & \text{if $\hat \alpha(v_r)> \hat \beta(v_l)$},\\
		|\hat\alpha(v_l)-\hat\alpha(v_r)| \hspace{-5pt } \mod 2   & \text{otherwise.}
	\end{cases}
	$
\end{Lemma}
\begin{proof}
	Suppose that $D\in \hat D_{\hat\alpha(v)}(v)$, $D_l = D \cap \hat{V}(v_l)$, and $D_r = D \cap \hat{V}(v_r)$. By Lemma~\ref{lemma:true min merge}, we have $D_l \in \Psi(v_l) \cap \hat D_s(v_l)$ and $D_r \in \Psi(v_r) \cap \hat D_t(v_r)$ for some $0 \le s \le |\hat{TS}(v_l)|$ and $0 \le t \le |\hat{TS}(v_r)|$. According to the induction hypothesis, Equation~(\ref{eq2}) holds for $v_l$ and $v_r$. Therefore, the possible index values of $s$ and $t$ are from $\hat \alpha(v_l)$ to $\hat\beta(v_l)$ and from $\hat\alpha(v_r)$ to $\hat\beta(v_r)$, respectively, and are increasing by $2$ at a time.  Since $v$ is labeled by true twin operation $\otimes$, the index value of $\hat\alpha(v)$ is determined by the two intervals $[\hat\alpha(v_l), \hat\beta(v_l)]$ and $[\hat\alpha(v_r),\hat\beta(v_r)]$.  More concretely, as the statement shows, we consider three cases that arise depending on how the two intervals overlap. The first two cases deal with the situation when the two intervals don't overlap. Meanwhile, the last case consider the situation when the two intervals overlap.	
\end{proof}

\begin{Lemma} \label{lemma:determine_beta_for_true}
	$\hat\beta(v)= \hat\beta(v_l)+\hat\beta(v_r)$.
\end{Lemma}
\begin{proof}
	We first show that $\hat\beta(v) \ge \hat\beta(v_l)+\hat\beta(v_r)$. Suppose $D_l \in \hat D_{\hat\beta(v_l)}(v_l)$ and $D_r \in \hat D_{\hat\beta(v_r)}(v_r)$ are two vertex sets of $G$ such that $G[D_l - X_l]$ and $G[D_r - X_r]$ have perfect matchings $M_l$ and $M_r$ with $|X_l| = \hat\beta(v_l)$ and $|X_r| = \hat\beta(v_r)$, respectively. Then, $D = D_l \cup D_r$ is a dominating set of $\hat{V}(v) - \hat{TS}(v)$ such that $G[D - (X_l \cup X_r)]$ contains a perfect matching $M_l \cup M_r$. Further, by Lemma \ref{lemma:true min merge}, we also have $D \in \Psi(v)$. Thus, $\hat\beta(v) \ge k = \hat\beta(v_l) +\hat\beta(v_r)$. 
	
	Next, we show that $\hat\beta(v) \le \hat\beta(v_l)+\hat\beta(v_r)$. Suppose that $D \in \hat D_{\hat\beta}(v)$, $D_l = D \cap \hat{V}(v_l)$, $D_r = D \cap \hat{V}(v_r)$, and $X \subseteq D \cap \hat{TS}(v)$ be a vertex set such that $G[D - X]$ has a perfect matching $M$. We also assume that $M$ contains exactly $h$ edges $(x_1, y_1), (x_2, y_2),\ldots, (x_h, y_h)$ such that $x_i \in \hat{TS}(v_l)$ and $y_i \in \hat{TS}(v_r)$ for $1 \le i \le h$. Moreover, the number of unpaired vertices in $\hat{TS}(v_l)$ and $\hat{TS}(v_r)$ are $k_l = |X \cap \hat{TS}(v_l)|$ and $k_r = |X \cap \hat{TS}(v_r)|$, respectively. Meanwhile, by Lemma \ref{lemma:true min merge}, we have $D_l \in \Psi(v_l) \cap \hat D_s(v_l)$ and $D_r \in \Psi(v_r) \cap \hat D_t(v_r)$ for some $0 \le s \le |\hat{TS}(v_l)|$ and $0 \le t \le |\hat{TS}(v_r)|$. Furthermore, it follows from Lemma~\ref{lemma:r_and_l_are_minimized_for_true} that $s = k_l + h$ and $t = k_r + h$. Thus, $\hat\beta(v) = k_l + k_r \le (k_l + h) + (k_r + h) = s + t \le \hat\beta(v_l) + \hat\beta(v_r)$.  
\end{proof}

\smallskip
Then we have the main result of Procedure~\ref{algo:determine_values_for_true}.

\begin{Lemma}\label{lemma:determine_values_for_true}
	Suppose that $v$ is annotated by true twin operation $\otimes$, then Procedure~\ref{algo:determine_values_for_true} determines $\hat{min}(v)$, $\hat\alpha(v)$, and $\hat\beta(v)$ in $O(1)$ time.
\end{Lemma}
\begin{proof}
	Combining Lemmas~\ref{lemma:determine_min_for_true},~\ref{lemma:determine_alpha_for_true}, and~\ref{lemma:determine_beta_for_true}, we obtain the correctness proof of the procedure. Moreover, the time complexity of the algorithm is $O(1)$ as all the steps can be easily implemented in $O(1)$ time.
\end{proof}

\smallskip
\subsubsection{Equation~(\ref{eq2}) holds for true twin operation $\otimes$} \label{subsection:Equation_holds_for_true_twin}
\ 
\vspace{9pt}
\newline
Suppose that $v$ is labeled by true twin operation $\otimes$ with left child $v_l$ and right child $v_r$. Below, we show that Equation~(\ref{eq2}) holds for $v$ according to the arguments mentioned in Corollary~\ref{corollary:proof_strategy}. For simplicity, let $min\text{-}ts = \min \setof{|\hat{TS}(v_l)|,|\hat{TS}(v_r)|}$.

\smallskip 
\begin{Lemma} \label{lemma:true gamma_0(G)=min {gamma_k(G_L) + gamma_k(G_R)}} 
	$\hat\gamma_0(v) = \min \setof{\hat\gamma_i(v_l)+\hat\gamma_i(v_r) \mid 0\le i \le min\text{-}ts}$.
\end{Lemma}
\begin{proof}
	Suppose that $D\in D_0(v)$, $D_l = D \cap \hat{V}(v_l)$, and $D_r = D \cap \hat{V}(v_r)$. We further assume $G[D]$ has a perfect matching $M$ containing $h$ edges $(x_1, y_1),$ $(x_2, y_2), \ldots,$ $(x_h, y_h)$ such that $x_i \in \hat{TS}(v_l)$ and $y_i \in \hat{TS}(v_r)$ for $1 \le i \le h$. According to Lemma~\ref{lemma:r_and_l_are_minimized_for_true}, we have $D_l \in \hat D_{h}(v_l)$ and $D_r \in \hat D_{h}(v_r)$. This implies that $\hat\gamma_0(v) = \hat\gamma_{h}(v_l) + \hat\gamma_{h}(v_r)$ for some $0 \le h \le min\text{-}ts$. Meanwhile, since we have examined each index to determine the minimum cardinality, the lemma certainly holds.
\end{proof}

\begin{Lemma} \label{lemma:true gamma_0(G)}
	$\hat\gamma_0(v)=\hat{min}(v)+\hat\alpha(v)$.
\end{Lemma}
\begin{proof}
	By Lemma~\ref{lemma:true gamma_0(G)=min {gamma_k(G_L) + gamma_k(G_R)}}, it suffices to show that $\min \{\hat\gamma_k(v_l) + \hat\gamma_k(v_r) \mid 0 \le k \le min\text{-}ts\} = \hat{min}(v)+\hat\alpha(v)$. Moreover, as described in Lemma~\ref{lemma:determine_alpha_for_true}, the value of $\hat\alpha(v)$ depends how the two intervals $[\hat\alpha(v_l),\hat\beta(v_l)]$ and $[\hat\alpha(v_r),\hat\beta(v_r)]$ overlap. Hence, we prove the statement by considering the three corresponding cases, and in each case we shall prove $\hat\gamma_k(v_l) + \hat\gamma_k(v_r)$ is minimized at $k = \min \setof{\hat\beta(v_l),\hat\beta(v_r)}$ and show $\hat\gamma_{k^*}(v_l)+\hat\gamma_{k^*}(v_r) = \hat{min}(v)+\hat\alpha(v)$ with $k^* = \min\{\hat\beta(v_l),\hat\beta(v_r)\}$. 
	
	First, we consider the case when $\hat\alpha(v_l) > \hat\beta(v_r)$. This implies that $k^* = \hat\beta(v_r)$ and $k^* < \hat\alpha(v_l)$. By the definition of $\hat\beta(v_r)$, we have $\hat\gamma_k(v_r) \ge \hat\gamma_{k^*}(v_r)$ for $k < k^*$. Moreover, by the inductive hypothesis, Equation~(\ref{eq2}) holds for $v_l$. Therefore, we have $\hat\gamma_k(v_l) > \hat\gamma_{k^*}(v_l)$ for $k < k^*$ as the slope of the equation is $-1$ when $k \le \hat\alpha(v)$. Hence, we have $\hat\gamma_k(v_l) + \hat\gamma_k(v_r) > \hat\gamma_{k^*}(v_l)+\hat\gamma_{k^*}(v_r)$ for $k < k^*$. Similarly, since Equation~(\ref{eq2}) also holds for $v_r$, when $k$ increases one unit, $\hat\gamma_k(v_r)$ increases one unit accordingly for $k^* < k  \le min\text{-}ts$. In addition, according to Lemma~\ref{lemma:k=k+-1}, when $k$ increases one unit, $\hat\gamma_k(v_l)$ decreases at most one unit for $k^* < k  \le min\text{-}ts$. It follows that $\hat\gamma_k(v_l) + \hat\gamma_k(v_r) \ge \hat\gamma_{k^*}(v_l)+\hat\gamma_{k^*}(v_r)$ for $k > k^*$. From the above discussion it can be seen that $\hat\gamma_k(v_l) + \hat\gamma_k(v_r)$ is minimized at $k = k^*$. Then, the lemma holds, as a consequence of

	\vspace{-14pt}
	\begin{eqnarray*}
		\hspace{-5pt}
		\hat\gamma_0(v) & = & \min \setof{\hat\gamma_k(v_l) + \hat\gamma_k(v_r) \mid 0 \le k \le min\text{-}ts} \text{~~~~(by Lemma~\ref{lemma:true gamma_0(G)=min {gamma_k(G_L) + gamma_k(G_R)}})}\\
		& = & \hat\gamma_{k^*}(v_l)+\hat\gamma_{k^*}(v_r)\\
		& = & \hat{min}(v_l) + \hat\alpha(v_l) - \hat\beta(v_r) +\hat{min}(v_r) \text{~~~~(by Equation~(\ref{eq2}))}\\
		& = & \hat{min}(v)+\hat\alpha(v). 
		\text{~~~~(by Lemmas~\ref{lemma:determine_min_for_true} and~\ref{lemma:determine_alpha_for_true}}) 
		\vspace{14pt}
	\end{eqnarray*}
	
	\noindent Symmetrically, one can prove the case when $\hat\alpha(v_r)>\hat\beta(v_l)$. 
	
	Finally, we consider the case when the two intervals overlap. Without loss of generality, we can assume that $k^* = \hat\beta(v_r)$. For the subcase when $\hat\alpha(v_l)-\hat\alpha(v_r)$ is even, one can see that $\hat\alpha(v_l)-\hat\gamma_{k^*}(v_l)$ is also even. Thus, by Equation~(\ref{eq2}), $\hat\gamma_{k^*}(v_l)=\hat{min}(v_l)$ and $\hat\gamma_{k^*}(v_r)=\hat{min}(v_r)$. Using similar arguments as above, we obtain $\hat\gamma_0(v) =  \hat\gamma_{k^*}(v_l) +\hat\gamma_{k^*}(v_r) = \hat{min}(v_l) + \hat{min}(v_r) = \hat{min}(v)+\hat\alpha(v)$. Moreover, for the subcase when $\hat\alpha(v_l)-\hat\alpha(v_r)$ is odd, one can show that $\hat\gamma_0(v) = \hat\gamma_{k^*}(v_l) +\hat\gamma_{k^*}(v_r) = \hat{min}(v_l) + \hat{min}(v_r) + 1 = \hat{min}(v)+\hat\alpha(v)$.  
\end{proof}

\begin{Lemma} \label{lemma:true gamma_ts(G)}
	$\hat\gamma_{|\hat{TS}(v)|}(v)=\hat{min}(v)+ |\hat{TS}(v)|-\hat\beta(v)$.
\end{Lemma}
\begin{proof}
	Suppose that $D\in \hat D_{|\hat{TS}(v)|}(v)$, $D_l = D \cap \hat{V}(v_l)$, $D_r = D \cap \hat{V}(v_r)$. By Lemma~\ref{lemma:r_and_l_are_minimized_for_true}, we have $D_l \in \hat D_{|\hat{TS}(v)|}(v_l)$ and $D_r \in \hat D_{|\hat{TS}(v_r)|}(v_r)$. Further, by the inductive hypothesis, Equation~(\ref{eq2}) holds for $v_l$ and $v_r$. Thus, the lemma holds, as a consequence of
	
	\vspace{-14pt}
	
	\begin{eqnarray*}
		\hat\gamma_{|\hat{TS}(v)|}(v) 
		& = & \hat\gamma_{|\hat{TS}(v_l)|}(v_l)+\hat\gamma_{|\hat{TS}(v_l)|}(v_r)\text{~~~~(by Lemma~\ref{lemma:r_and_l_are_minimized_for_true})}\\
		& = & \hat{min}(v_l)+ |\hat{TS}(v_l)|-\hat\beta(v_l) +  \hat{min}(v_r)+|\hat{TS}(v_r)|-\hat\beta(v_r)\text{~~~~(by Equation~(\ref{eq2}))}\\
		& = & \hat{min}(v)+ |\hat{TS}(v)|-\hat\beta(v). 
		\text{~~~~(by Lemmas~\ref{lemma:determine_min_for_true} and~\ref{lemma:determine_beta_for_true}})
	\end{eqnarray*}
	
	\vspace{-16pt}
\end{proof}

\vspace{0pt}
\begin{Lemma} \label{lemma:true gamma_alpha+2i(G)=gamma_min(G)}
	$\hat\gamma_{\hat\alpha(v)+2i}(v) = \hat{min}(v)$ for $0 \le i \le (\hat\beta(v)-\hat\alpha(v))/2$.
\end{Lemma}
\begin{proof} 
	We prove the statement by induction on $i$ from 0 to $(\hat\beta(v)-\hat\alpha(v))/2$. The assertion certainly holds for $i = 0$ as $\hat\gamma_{\hat\alpha(v)}(v) = \hat{min}(v)$. Assume $\hat\gamma_{\hat\alpha(v)+2i}(v) = \hat{min}(v)$, to show $\hat\gamma_{\hat\alpha(v)+ 2(i + 1)}(v) = \hat{min}(v)$, we only need to find a vertex set $S \in \hat D_{\hat\alpha(v)+2(i+1)}(v)$ and verify that $\hat D_{\hat\alpha(v)+2(i+1)}(v) \subseteq \Psi(v)$. Suppose $D \in \hat D_{\hat\alpha(v)+2i}(v)$ and $X \subseteq D \cap \hat{TS}(v)$ is a vertex set such that $G[D - X]$ has a perfect matching $M$ with $|X| = \hat\alpha(v)+ 2i$.  If $M$ contains an edge $(x, y)$ such that $x \in \hat{TS}(v_l)$ and $y \in \hat{TS}(v_r)$, then $X \cup \setof{x, y}$ is a set of unpaired vertex with respect to $(D,M - \setof{(x, y)})$. Moreover, since $D \in \hat D_{\hat\alpha(v)+2i}(v)$ and $\hat D_{\hat\alpha(v)+2i}(v) \subseteq \Psi(v)$, one can see that $D \in \hat D_{\hat\alpha(v)+2(i+1)}(v)$ and so $\hat D_{\hat\alpha(v)+2(i+1)}(v) \subseteq \Psi(v)$. Consequently, $\hat\gamma_{\hat\alpha(v)+ 2(i + 1)}(v) = \hat{min}(v)$.

	Below, we consider the case when $M$ does not contain the edge $(x,y)$. Suppose that  $D_l = D\cap \hat{V}(v_l)$, $D_r = D \cap \hat{V}(v_r)$, $k_l = |X \cap \hat{TS}(v_l)|$ and $k_r = |X \cap \hat{TS}(v_r)|$. According to Lemmas~\ref{lemma:true min merge} and~\ref{lemma:determine_beta_for_true}, we have $D_l \in \Psi(v_l)$, $D_r \in \Psi(v_r)$ and $\hat\beta(v) = \hat\beta(v_l) + \hat\beta(v_r)$, respectively. Moreover, since $k_l + k_r = \hat\alpha(v)+ 2i < \hat \beta(v)$, either $k_l<\hat\beta(v_l)$ or $k_r<\hat\beta(v_r)$. Without loss of generality, we can assume that $k_l<\hat\beta(v_l)$. Now that Equation~(\ref{eq2}) holds for $v_l$ by the inductive hypothesis, there exists a vertex set $D'_l \in \hat D_{k_l+2}(v_l)$ and $\hat D_{k_l+2}(v_l) \subseteq\Psi(v_l)$. By Lemma \ref{lemma:true min merge}, we can construct a vertex set $S = D'_l \cup D_r$ such that $S \in\Psi(v)$. Furthermore, since $D'_l \in \hat D_{k_l+2}(v_l)$, $D_r \in \hat D_{k_r}(v_r)$, and $k_l + k_r = \alpha(v) + 2i$, one can verify that $S \in \hat D_{\hat\alpha(v)+2(i+1)}(v)$ and so $\hat D_{\hat\alpha(v)+2(i+1)}(v) \subseteq \Psi(v)$. It follows that $\hat\gamma_{\hat\alpha(v)+ 2(i + 1)}(v) = \hat{min}(v)$.
\end{proof}

\smallskip
Combining Corollary \ref{corollary:proof_strategy} and Lemmas~\ref{lemma:true gamma_0(G)},~\ref{lemma:true gamma_ts(G)}, and \ref{lemma:true gamma_alpha+2i(G)=gamma_min(G)}, we obtain the following main result of this subsection.

\vspace{6pt}
\begin{Lemma} \label{lemma:eq2_holds_for_true}
	Suppose that $v$ is labeled by true twin operation $\otimes$ with left child $v_l$ and right child $v_r$. If Equation~(\ref{eq2}) holds for $v_l$ and $v_r$, then the equation also holds for $v$.
\end{Lemma}

\medskip
\subsection{False Twin Operation $(G=G_l \odot G_r)$} 
\label{subsection:false_twin}
In this subsection, we assume that an internal vertex $v$ of $T$ is labeled by false twin operation $\odot$ with left child $v_l$ and right child $v_r$. A procedure, which was developed for determining $\hat{min}(v)$, $\hat\alpha(v)$, and $\hat\beta(v)$, is first presented. Then, we show that Equation~(\ref{eq2}) holds for $v$ by applying the strategy mentioned in Corollary~\ref{corollary:proof_strategy}.

\subsubsection{Determine $\hat{min}(v)$, $\hat\alpha(v)$, and $\hat\beta(v)$ for false twin operation} \label{subsection:determine_domination-number_for_false_twin}
\ 
\vspace{9pt}
\newline
Below, we first propose a procedure for finding $\hat{min}(v)$, $\hat\alpha(v)$, and $\hat\beta(v)$ and then present lemmas for proving the correctness. Each step of the procedure is simple and easy to implement, but the proof is more complicated.

\vspace{-6pt}
\floatname{algorithm}{Procedure}
\begin{algorithm}[htb]
	\caption{Determine the values $\hat{min}(v)$, $\hat\alpha(v)$, and $\hat\beta(v)$ for false twin operation $\odot$}
	\label{algo:determine_values_for_false}
	\begin{algorithmic} [1]
		\baselineskip 14pt
		\REQUIRE $\hat{min}(v_i)$, $\hat\alpha(v_i)$, and $\hat\beta(v_i)$ with $i \in \setof{l,r}$.
		
		\ENSURE $\hat{min}(v)$, $\hat\alpha(v)$, and $\hat\beta(v)$.
		
		\STATE let $\hat{min}(v) \leftarrow \hat{min}(v_l) + \hat{min}(v_r)$;
		
		\STATE let $\hat\alpha(v) \leftarrow \hat\alpha(v_l) + \hat\alpha(v_r)$;
		
		\STATE let $\hat\beta(v) \leftarrow \hat \beta(v_l) + \hat \beta(v_r)$;
		
		\RETURN $\hat{min}(v)$, $\hat\alpha(v)$, and $\hat\beta(v)$.
	\end{algorithmic}
\end{algorithm} \baselineskip 14pt

\vspace{-6pt}
Suppose that $D\in \hat D_k(v)$ and $X \subseteq D \cap \hat{TS}(v)$ is a vertex set such that $G[D - X]$ has a perfect matching $M$ with $|X| = k$. The number of unpaired vertices in $\hat{TS}(v_l)$ and $\hat{TS}(v_r)$ are $k_l = |X \cap \hat{TS}(v_l)|$ and $k_r = |X \cap \hat{TS}(v_r)|$, respectively. Meanwhile, we set $M_l = M \cap \hat{E}(v_l)$, $M_r = M \cap \hat{E}(v_r)$, $X_l = (X \cap \hat{TS}(v_l))$, and $X_r = (X \cap \hat{TS}(v_r))$. Noticed that there is no edge $(x,y) \in M$ such that $x\in\hat{TS}(v_l)$ and $y\in\hat{TS}(v_r)$.

\begin{Lemma} \label{lemma:r_and_l_are_minimized_for_false}
	Suppose that $D\in \hat D_k(v)$, $D_l = D \cap \hat{V}(v_l)$, and $D_r = D \cap \hat{V}(v_r)$. Then, $D_l \in \hat D_{k_l}(v_l)$ and $D_r \in \hat D_{k_r}(v_r)$. Furthermore, $X_l$ and $X_r$ are two sets of unpaired vertices with respect to $(D_l, M_l)$ and $(D_r, M_r)$, respectively.
\end{Lemma}
\begin{proof}
	Clearly, $G[D_l - X_l]$ contains a perfect matching $M_l$ with $|X_l| = k_l$. Since there exists no edge $(x, y) \in \hat E(v)$ such that $x \in \hat{V}(v_l) - \hat{TS}(v_l)$ and $y \in D_r$, $D_l$ is a dominating set of  $\hat{V}(v_l)-\hat{TS}(v_l)$. Therefore, to prove $D_l \in \hat D_{k_l}(v_l)$, it suffices to show that the number of vertices in $D_l$ is minimized. Suppose for the purpose of contradiction that $S_l \in \hat D_{k_l}(v_l)$ and $|S_l| < |D_l|$. There is a set of unpaired vertices $X'_l$ such that $G[S_l-X'_l]$ contains a perfect matching with $|X'_l|=k_l$. Then, one can verify that $S_l \cup D_r$ is also a dominating set of  $\hat{V}(v)-\hat{TS}(v)$. Further, since each vertex between $\hat{TS}(v_l)$ and $\hat{TS}(v_r)$ is independent, we can construct a set of unpaired vertices $X'$ as an union of $X'_l$ and $X_r$ such that $G[\{S_l \cup D_r\}-X']$ has a perfect matching and $|X'|=k$. This contradicts the fact $D\in \hat D_k(v)$ as $|S_l| + |D_r| < |D|$. Symmetrically, it can be shown that $D_r \in \hat D_{k_r}(v_r)$ and $G[D_r - X_r]$ contains a perfect matching $M_r$ with $|X_r| = k_r$.
\end{proof}

\begin{Lemma} \label{lemma:determine_min_for_false}
	$\hat{min}(v)=\hat{min}(v_l)+\hat{min}(v_r)$.
\end{Lemma}
\begin{proof}
	We first show that $\hat{min}(v) \ge \hat{min}(v_l)+\hat{min}(v_r)$. Suppose that $D\in \Psi(v)$, $D_l = D \cap \hat{V}(v_l)$, and $D_r = D \cap \hat{V}(v_r)$. Then, by Lemma~\ref{lemma:r_and_l_are_minimized_for_false}, we have $D_l \in \hat D_s(v_l)$ and $D_r \in \hat D_t(v_r)$ for some $0 \le s \le |\hat{TS}(v_l)|$ and $0 \le t \le |\hat{TS}(v_r)|$. Hence, $|D_l|= \hat\gamma_{s}(v_l) \ge \hat{min}(v_l)$ and $|D_r|= \hat\gamma_{t}(v_r) \ge \hat{min}(v_r)$. It follows that $\hat{min}(v) = |D| = |D_l|+ |D_r| \ge \hat{min}(v_l)+\hat{min}(v_r)$. Next, we show that $\hat{min}(v) \le \hat{min}(v_l)+\hat{min}(v_r)$. Suppose that $D_l \in \Psi(v_l)$ and $D_r \in \Psi(v_r)$ are two vertex sets such that $G[D_l - X_l]$ and $G[D_r - X_r]$ have perfect matchings $M_l$ and $M_r$, respectively. Then, $D = D_l \cup D_r$ is dominating set of $\hat{V}(v) - \hat{TS}(v)$ such that $G[D - (X_l \cup X_r)]$ has a perfect matchings $M_l \cup M_r$. Hence, we have $\hat{min}(v) \le |D| = \hat{min}(v_l) + \hat{min}(v_r)$.
\end{proof}

\smallskip
Before proceeding to prove the correctness of $\hat\alpha(v)$ and $\hat\beta(v)$, we require the following auxiliary lemma. 

\begin{Lemma} \label{lemma:false min merge}
	Suppose that $D\in \hat{V}(v)$, $D_l = D \cap \hat{V}(v_l)$ and $D_r = D \cap \hat{V}(v_r)$. Then, 
	we have $D\in \Psi(v)$ if and only if $D_l \in \Psi(v_l)$ and $D_r \in \Psi(v_r)$.
\end{Lemma}
\begin{proof} 
	First we show the sufficient condition. Suppose that $D\in \Psi(v)$. Then, by Lemma~\ref{lemma:r_and_l_are_minimized_for_false}, we have $D_l \in \hat D_s(v_l)$ and $D_r \in \hat D_t(v_r)$ for some $0 \le s \le |\hat{TS}(v_l)|$ and $0 \le t \le |\hat{TS}(v_r)|$. Hence, it remains to show $\hat D_s(v_l) \subseteq \Psi(v_l)$ and $\hat D_t(v_r) \subseteq \Psi(v_r)$. By Lemma~\ref{lemma:determine_min_for_false}, we have $|D_l| + |D_r| = |D| = \hat{min}(v_l) + \hat{min}(v_r)$. In addition, since $|D_l|\ge \hat{min}(v_l)$ and $|D_r|\ge \hat{min}(v_r)$, we have $|D_l| = \hat{min}(v_l)$ and $|D_r| = \hat{min}(v_r)$. This implies that $\hat D_s(v_l) \subseteq \Psi(v_l)$ and $\hat D_t(v_r) \subseteq \Psi(v_r)$.
	
	Let us go to the necessity. Suppose that $D_l \in \Psi(v_l) \cap \hat D_s(v_l)$ and $D_r \in \Psi(v_r) \cap \hat D_s(v_r)$ are two vertex sets such that $G[D_l - X_l]$ and $G[D_r - X_r]$ have perfect matchings $M_l$ and $M_r$ with $|X_l| = s$ and $|X_r| = t$, respectively. Then, $D$ is dominating set of $\hat{V}(v) - \hat{TS}(v)$ such that $G[D - (X_l \cup X_r)]$ has a perfect matchings $M_l \cup M_r$ clearly. Further, because of $|D| = |D_l| + |D_r| = \hat{min}(v_l) + \hat{min}(v_r) = \hat{min}(v)$, we have $D\in \Psi(v)$.
\end{proof}

\begin{Lemma} \label{lemma:determine_alpha_for_false}
	$\hat\alpha(v) =\hat \alpha(v_l) + \hat \alpha(v_r)$
\end{Lemma}
\begin{proof}
	We first show that $\hat\alpha(v) \le \hat\alpha(v_l)+\hat\alpha(v_r)$. Suppose $D_l \in \hat D_{\hat\alpha(v_l)}(v_l)$ and $D_r \in \hat D_{\hat\alpha(v_r)}(v_r)$ are two vertex sets of $G$ such that $G[D_l - X_l]$ and $G[D_r - X_r]$ have perfect matchings $M_l$ and $M_r$ with $|X_l| = \hat\alpha(v_l)$ and $|X_r| = \hat\alpha(v_r)$, respectively. Then, $D = D_l \cup D_r$ is a dominating set of $\hat{V}(v) - \hat{TS}(v)$ such that $G[D - (X_l \cup X_r)]$ contains a perfect matching $M_l \cup M_r$. It follows that $D \in \hat D_k(v)$ with $k = \hat\alpha(v_l) + \hat\alpha(v_r)$. Further, by Lemma \ref{lemma:false min merge}, we also have $D \in \Psi(v)$. Thus, $\hat\alpha(v) \le k = \hat\alpha(v_l) +\hat\alpha(v_r)$. 
	
	Next, we show that $\hat\alpha(v) \ge \hat\alpha(v_l)+\hat\alpha(v_r)$. Suppose that $D \in \hat D_{\hat\alpha}(v_l)$, $D_l = D \cap \hat{V}(v_l)$, $D_r = D \cap \hat{V}(v_r)$, and $X \subseteq D \cap \hat{TS}(v)$ be a vertex set such that $G[D - X]$ has a perfect matching $M$. Moreover, we assume the number of unpaired vertices in $\hat{TS}(v_l)$ and $\hat{TS}(v_r)$ are $k_l = |X \cap \hat{TS}(v_l)|$ and $k_r = |X \cap \hat{TS}(v_r)|$, respectively. Meanwhile, by Lemma \ref{lemma:false min merge}, we have $D_l \in \Psi(v_l) \cap \hat D_s(v_l)$ and $D_r \in \Psi(v_r) \cap \hat D_t(v_r)$ for some $0 \le s \le |\hat{TS}(v_l)|$ and $0 \le t \le |\hat{TS}(v_r)|$. Furthermore, it follows from Lemma~\ref{lemma:r_and_l_are_minimized_for_false} that $s = k_l$ and $t = k_r$. Thus, $\hat\alpha(v) = k_l + k_r = s + t \ge \hat\alpha(v_l) + \hat\alpha(v_r)$.  
\end{proof}

\begin{Lemma} \label{lemma:determine_beta_for_false}
	$\hat\beta(v)= \hat\beta(v_l)+\hat\beta(v_r)$.
\end{Lemma}
\begin{proof}
	We first show that $\hat\beta(v) \ge \hat\beta(v_l)+\hat\beta(v_r)$. Suppose $D_l \in \hat D_{\hat\beta(v_l)}(v_l)$ and $D_r \in \hat D_{\hat\beta(v_r)}(v_r)$ are two vertex sets of $G$ such that $G[D_l - X_l]$ and $G[D_r - X_r]$ have perfect matchings $M_l$ and $M_r$ with $|X_l| = \hat\beta(v_l)$ and $|X_r| = \hat\beta(v_r)$, respectively. Then, $D = D_l \cup D_r$ is a dominating set of $\hat{V}(v) - \hat{TS}(v)$ such that $G[D - (X_l \cup X_r)]$ contains a perfect matching $M_l \cup M_r$. It follows that $D \in \hat D_k(v)$ with $k = \hat\beta(v_l) + \hat\beta(v_r)$. Further, by Lemma \ref{lemma:false min merge}, we also have $D \in \Psi(v)$. Thus, $\hat\beta(v) \ge k = \hat\beta(v_l) +\hat\beta(v_r)$. 
	
	Next, we show that $\hat\beta(v) \le \hat\beta(v_l)+\hat\beta(v_r)$. Suppose that $D \in \hat D_{\hat\beta}(v_l)$, $D_l = D \cap \hat{V}(v_l)$, $D_r = D \cap \hat{V}(v_r)$, and $X \subseteq D \cap \hat{TS}(v)$ be a vertex set such that $G[D - X]$ has a perfect matching $M$. We also assume that $M$ contains exactly $h$ edges $(x_1, y_1), (x_2, y_2),\ldots, (x_h, y_h)$ such that $x_i \in \hat{TS}(v_l)$ and $y_i \in \hat{TS}(v_r)$ for $1 \le i \le h$. Moreover, we assume the number of unpaired vertices in $\hat{TS}(v_l)$ and $\hat{TS}(v_r)$ are $k_l = |X \cap \hat{TS}(v_l)|$ and $k_r = |X \cap \hat{TS}(v_r)|$, respectively. Meanwhile, by Lemma \ref{lemma:false min merge}, we have $D_l \in \Psi(v_l) \cap \hat D_s(v_l)$ and $D_r \in \Psi(v_r) \cap \hat D_t(v_r)$ for some $0 \le s \le |\hat{TS}(v_l)|$ and $0 \le t \le |\hat{TS}(v_r)|$. Furthermore, it follows from Lemma~\ref{lemma:r_and_l_are_minimized_for_false} that $s = k_l$ and $t = k_r$. Thus, $\hat\beta(v) = k_l + k_r = s + t \le \hat\beta(v_l) + \hat\beta(v_r)$.  
\end{proof}

\smallskip
Then we have the main result of Procedure~\ref{algo:determine_values_for_false}.

\begin{Lemma}\label{lemma:determine_values_for_false}
	Suppose that $v$ is annotated by false twin operation $\odot$, then Procedure~\ref{algo:determine_values_for_false} determines $\hat{min}(v)$, $\hat\alpha(v)$, and $\hat\beta(v)$ in $O(1)$ time.
\end{Lemma}
\begin{proof}
	Combining Lemmas~\ref{lemma:determine_min_for_false},~\ref{lemma:determine_alpha_for_false}, and~\ref{lemma:determine_beta_for_false}, we obtain the correctness proof of the procedure. Moreover, the time complexity of the algorithm is $O(1)$ as all the steps can be easily implemented in $O(1)$ time.
\end{proof}

\smallskip
\subsubsection{Equation~(\ref{eq2}) holds for false twin operation $\odot$} \label{subsection:Equation_holds_for_false_twin}
\ 
\vspace{9pt}
\newline
Suppose that $v$ is labeled by false twin operation $\otimes$ with left child $v_l$ and right child $v_r$. Below, we show that Equation~(\ref{eq2}) holds for $v$ according to the arguments mentioned in Corollary~\ref{corollary:proof_strategy}. For simplicity, let $min\text{-}ts = \min \setof{|\hat{TS}(v_l)|,|\hat{TS}(v_r)|}$.

\begin{Lemma} \label{lemma:false gamma_0(G)}
	$\hat\gamma_0(v)=\hat{min}(v)+\hat\alpha(v)$.
\end{Lemma}
\begin{proof}
	Suppose that $D\in D_0(v)$, $D_l = D \cap \hat{V}(v_l)$ and $D_r = D \cap \hat{V}(v_r)$. By Lemma~\ref{lemma:r_and_l_are_minimized_for_false}, we have $D_l \in \hat D_{0}(v_l)$ and $D_r \in \hat D_{0}(v_r)$. Further, by the inductive hypothesis, Equation~(\ref{eq2}) holds for $v_l$ and $v_r$. Thus, the lemma holds, as a consequence of
	
	\vspace{-20pt}
	
	\begin{eqnarray*}
		\hat\gamma_0(v) 
		& = & \hat\gamma_0(v_l)+\hat\gamma_0(v_r)\text{~~~~(by Lemma~\ref{lemma:r_and_l_are_minimized_for_false})}\\
		& = & \hat{min}(v_l)+ \hat\alpha(v_l) +  \hat{min}(v_r)+ \hat\alpha(v_r)\text{~~~~(by Equation~(\ref{eq2}))}\\
		& = & \hat{min}(v)+ \hat\alpha(v). 
		\text{~~~~(by Lemmas~\ref{lemma:determine_min_for_false} and~\ref{lemma:determine_beta_for_false}})
	\end{eqnarray*}
	
	\vspace{-6pt}
\end{proof}

\begin{Lemma} \label{lemma:false gamma_ts(G)}
	$\hat\gamma_{|\hat{TS}(v)|}(v)=\hat{min}(v)+ |\hat{TS}(v)|-\hat\beta(v)$.
\end{Lemma}
\begin{proof}
	Suppose that $D\in \hat D_{|\hat{TS}(v)|}(v)$, $D_l = D \cap \hat{V}(v_l)$, $D_r = D \cap \hat{V}(v_r)$. By Lemma~\ref{lemma:r_and_l_are_minimized_for_false}, we have $D_l \in \hat D_{|\hat{TS}(v)|}(v_l)$ and $D_r \in \hat D_{|\hat{TS}(v_r)|}(v_r)$. Further, by the inductive hypothesis, Equation~(\ref{eq2}) holds for $v_l$ and $v_r$. Thus, the lemma holds, as a consequence of
	
	\vspace{-14pt}
	
	\begin{eqnarray*}
		\hat\gamma_{|\hat{TS}(v)|}(v) 
		& = & \hat\gamma_{|\hat{TS}(v_l)|}(v_l)+\hat\gamma_{|\hat{TS}(v_l)|}(v_r)\text{~~~~(by Lemma~\ref{lemma:r_and_l_are_minimized_for_false})}\\
		& = & \hat{min}(v_l)+ |\hat{TS}(v_l)|-\hat\beta(v_l) +  \hat{min}(v_r)+|\hat{TS}(v_r)|-\hat\beta(v_r)\text{~~~~(by Equation~(\ref{eq2}))}\\
		& = & \hat{min}(v)+ |\hat{TS}(v)|-\hat\beta(v). 
		\text{~~~~(by Lemmas~\ref{lemma:determine_min_for_false} and~\ref{lemma:determine_beta_for_false}})
	\end{eqnarray*}
	
	\vspace{-20pt}
\end{proof}

\smallskip
\begin{Lemma} \label{lemma:false gamma_alpha+2i(G)=gamma_min(G)}
	$\hat\gamma_{\hat\alpha(v)+2i}(v) = \hat{min}(v)$ for $0 \le i \le (\hat\beta(v)-\hat\alpha(v))/2$.
\end{Lemma}
\begin{proof}
	We prove the statement by induction on $i$ from 0 to $(\hat\beta(v)-\hat\alpha(v))/2$. The assertion certainly holds for $i = 0$ as $\hat\gamma_{\hat\alpha(v)}(v) = \hat{min}(v)$. Assume $\hat\gamma_{\hat\alpha(v)+2i}(v) = \hat{min}(v)$, to show $\hat\gamma_{\hat\alpha(v)+ 2(i + 1)}(v) = \hat{min}(v)$, we only need to find a vertex set $S \in \hat D_{\hat\alpha(v)+2(i+1)}(v)$ and verify that $\hat D_{\hat\alpha(v)+2(i+1)}(v) \subseteq \Psi(v)$. Suppose $D \in \hat D_{\hat\alpha(v)+2i}(v)$ and $X \subseteq D \cap \hat{TS}(v)$ is a vertex set such that $G[D - X]$ has a perfect matching $M$. We set  $D_l = D\cap \hat{V}(v_l)$, $D_r = D \cap \hat{V}(v_r)$, $k_l = |X \cap \hat{TS}(v_l)|$ and $k_r = |X \cap \hat{TS}(v_r)|$. According to Lemmas~\ref{lemma:false min merge} and~\ref{lemma:determine_beta_for_false}, we have $D_l \in \Psi(v_l)$, $D_r \in \Psi(v_r)$ and $\hat\beta(v) = \hat\beta(v_l) + \hat\beta(v_r)$, respectively. Moreover, since $k_l + k_r = \hat\alpha(v)+ 2i < \hat \beta(v)$, either $k_l<\hat\beta(v_l)$ or $k_r<\hat\beta(v_r)$. Without loss of generality, we can assume that $k_l<\hat\beta(v_l)$. Now that Equation~(\ref{eq2}) holds for $v_l$ by the inductive hypothesis, there exists a vertex set $D'_l \in \hat D_{k_l+2}$ and $\hat D_{k_l+2} \subseteq\Psi(v_l)$. By Lemma \ref{lemma:false min merge}, we can construct a vertex set $S = D'_l \cup D_r$ such that $S \in\Psi(v)$. Furthermore, since $D'_l \in \hat D_{k_l+2}(v)$, $D_r \in \hat D_{k_r}(v)$, and $k_l + k_r = \alpha(v) + 2i$, one can verify that $S \in \hat D_{\hat\alpha(v)+2(i+1)}(v)$ and so $\hat D_{\hat\alpha(v)+2(i+1)}(v) \subseteq \Psi(v)$. It follows that $\hat\gamma_{\hat\alpha(v)+ 2(i + 1)}(v) = \hat{min}(v)$.
\end{proof}

\smallskip
Combining Corollary \ref{corollary:proof_strategy} and Lemmas~\ref{lemma:false gamma_0(G)},~\ref{lemma:false gamma_ts(G)}, and \ref{lemma:false gamma_alpha+2i(G)=gamma_min(G)}, we obtain the following main result of this subsection.

\begin{Lemma} \label{lemma:eq2_holds_for_false}
	Suppose that $v$ is labeled by false twin operation $\odot$ with left child $v_l$ and right child $v_r$. If Equation~(\ref{eq2}) holds for $v_l$ and $v_r$, then the equation also holds for $v$.
\end{Lemma}

\medskip
\subsection{Attachment Operation $(G=G_l \oplus G_r)$} 
\label{subsection:attachment}
In this subsection, we assume that an internal vertex $v$ of $T$ is labeled by attachment operation $\oplus$ with left child $v_l$ and right child $v_r$. Recall that the attachment operation has the same vertex set and edge set as the true twin operation, i.e., $\hat{V}(v)=\hat{V}(v_l)\cup \hat{V}(v_r)$ and $\hat{E}(v)=\hat{E}(v_l)\cup \hat{E}(v_r)\cup \{(x,y) \mid  x\in \hat{TS}(v_l) ~\text{and}~  y\in \hat{TS}(v_r)\}$. Therefore, the discussion for attachment operation and true twin operation are highly similar. However, unlike the true twin operation has twin set $\hat{TS}(v) = \hat{TS}(v_l) \cup \hat{TS}(v_r)$, the attachment operation has only $\hat{TS}(v) = \hat{TS}(v_l)$. Thus, we need to ensure that the vertices in $\hat{TS}(v_r)$ are dominated and the vertices of dominating set in $\hat{TS}(v_r)$ are paired. In order to do so, we solve the problem by considering three cases $C_1, C_2$, and $C_3$, all of which are dependent on the two conditions $D_1$ and $D_2$:
\vspace{10pt}
\\
\mbox{} \hspace{55pt} $D_1$ : $\hat\alpha(v_r)>\hat\beta(v_l)$; \\
\mbox{} \hspace{55pt} $D_2$ : $\hat\alpha(v_l) = \hat\beta(v_r) =0$ or $\hat\alpha(v_r) = \hat\beta(v_l) = 0$.  
\vspace{10pt}

\noindent For $1 \le i \le 3$, the case $C_i = (d_1, d_2)$ is an ordered pair. If condition $D_j$ holds, then $d_j = 1$; and $d_j = 0$ otherwise, for $1 \le j \le 2$. We define the cases $C_1 = (0,0)$,  $C_2 = (1,0)$, and $C_3 = (0,1)$. Since conditions $D_1$ and $D_2$ are mutually exclusive, one can verify that all the possible cases have been considered. Moreover, for ease of subsequent discussion, we consider the three cases $C_1, C_2$, and $C_3$, respectively, in Subsections~\ref{subsubsection:C_1}--\ref{subsubsection:C_3}. In each of the Subsections, we first propose a procedure for determining $\hat{min}(v)$, $\hat\alpha(v)$, and $\hat\beta(v)$. Then, we shall show that Equation~(\ref{eq2}) holds for $v$.

Before that, some notations and useful properties are introduced below. Suppose that $D\in \hat D_k(v)$ and $X \subseteq D \cap \hat{TS}(v)$ is a vertex set such that $G[D - X]$ has a perfect matching $M$ with $|X| = k$. Suppose further that $M$ contains exactly $h$ edges $(x_1, y_1), (x_2, y_2),\ldots, (x_h, y_h)$ such that $x_i \in \hat{TS}(v_l)$ and $y_i \in \hat{TS}(v_r)$ for $1 \le i \le h$. Meanwhile, we set $M_l = M \cap \hat{E}(v_l)$, $M_r = M \cap \hat{E}(v_r)$, $X_l = X \cup \setof{x_1,x_2,\ldots,x_h}$, and $X_r =  \setof{y_1,y_2,\ldots,y_h}$.

\begin{Lemma} \label{lemma:r_and_l_are_minimized_for_attach}
	Suppose that $D\in \hat D_k(v)$, $D_l = D \cap \hat{V}(v_l)$ and $D_r = D \cap \hat{V}(v_r)$. If $X_l \not = \emptyset$, then $D_l \in \hat D_{k + h}(v_l)$ and $D_r \in \hat D_{h}(v_r)$. Furthermore, $X_l$ and $X_r$ are two sets of unpaired vertices with respect to $(D_l, M_l)$ and $(D_r, M_r)$, respectively.
\end{Lemma}
\begin{proof}
	Since $X_l \not = \emptyset$, the vertice in $\hat{TS}(v_r)$ are dominated. For the remaining part of the proof, one can show that the correctness holds by using a similar method of the arguments in Lemma~\ref{lemma:r_and_l_are_minimized_for_true}. So we omit the details.
\end{proof}

\smallskip
The next lemma describes relationships between attachment operation $\oplus$ and true twin operation $\otimes$, which will be helpful in determining $\hat{min}(v)$, $\hat\alpha(v)$, and $\hat\beta(v)$. To simplify description, we use $\ddot{v}$ to denote an internal vertex labeled by true twin operation~$\otimes$ with left child $v_l$ and right child $v_r$. 

\begin{Lemma} \label{lemma:attach >= true for gamma_k}
	$\hat\gamma_k(v)\ge \hat\gamma_k(\ddot{v})$ for $0\leq k\leq |\hat{TS}(v)|$. 
\end{Lemma}
\begin{proof}
	Let $D\in \hat D_k(v)$. Suppose $X \subseteq D \cap \hat{TS}(v)$ is a vertex set such that $G[D - X]$ has a perfect matching $M$ with $|X| = k$. Since $\hat{TS}(v)\subseteq \hat{TS}(\ddot{v})$, one can verify that $\hat{V}(\ddot{v})-\hat{TS}(\ddot{v})\subseteq N_{\hat{G}(\ddot{v})}[D]$ and $X \subseteq D \cap \hat{TS}(\ddot{v})$ is a set of unpaired vertices with respect to $(D, M)$. These imply that $\hat\gamma_k(v)\geq\hat\gamma_k(\ddot{v})$.
\end{proof}

\begin{Lemma} \label{lemma:attach >= true_for_min}
	$\hat{min}(v)\geq \hat{min}(v_l)+\hat{min}(v_r)$.
\end{Lemma}
\begin{proof}
	The assertion holds, as a consequence of	
	\vspace{-18pt}
	
	\begin{eqnarray*}
		\hat{min}(v) 
		& = & \min \setof{\hat\gamma_k(v)\mid 0\le k\le |\hat{TS}(v)|}\\
		& \ge & \min \setof{\hat\gamma_k(\ddot{v})\mid 0\le k\le |\hat{TS}(\ddot{v})|} \text{~~~~(by Lemma~\ref{lemma:attach >= true for gamma_k})}\\
		& = & \hat{min}(\ddot{v})\\
		& = & \hat{min}(v_l)+\hat{min}(v_r). 
		\text{~~~~(by Lemma~\ref{lemma:determine_min_for_true})}
	\end{eqnarray*}
	
	\vspace{-18pt}
\end{proof}


\smallskip
\subsubsection{Case $C_1$: conditions $D_1$ and $D_2$ are false} 
\label{subsubsection:C_1}
\ 
\vspace{9pt}
\newline
In this subsection, we first propose a procedure of determining $\hat{min}(v)$, $\hat\alpha(v)$, and $\hat\beta(v)$. As you will see, the procedure is highly similar to the one for true twin operation. After describing the procedure, we shall show that Equation~(\ref{eq2}) is true.

\paragraph{Determine $\hat{min}(v)$, $\hat\alpha(v)$, and $\hat\beta(v)$ for case $C_1$} \label{subsection:determine_domination-number_for_case_C_1}

\vspace{-16pt}
\floatname{algorithm}{Procedure}
\begin{algorithm}[htb]
	\caption{Determine the values $\hat{min}(v)$, $\hat\alpha(v)$, and $\hat\beta(v)$ for Case $C_1$}
	\label{algo:determine_values_for_attach_g}
	\begin{algorithmic} [1]
		\baselineskip 14pt
		\REQUIRE $\hat{min}(v_i)$, $\hat\alpha(v_i)$, and $\hat\beta(v_i)$ with $i \in \setof{l,r}$.
		
		\ENSURE $\hat{min}(v)$, $\hat\alpha(v)$, and $\hat\beta(v)$.
		
		\STATE let $\hat{min}(v) \leftarrow \hat{min}(v_l) + \hat{min}(v_r)$;
		
		\STATE let $\hat\alpha(v) \leftarrow \max \setof{\hat\alpha(v_l)-\hat\beta(v_r), |\hat\alpha(v_l)-\hat\alpha(v_r)| \hspace{-5pt }\mod 2}$;
		
		\STATE let $\hat\beta(v) \leftarrow \hat\beta(v_l)-\hat\alpha(v_r)$;
		
		\RETURN $\hat{min}(v)$, $\hat\alpha(v)$, and $\hat\beta(v)$.
	\end{algorithmic}
\end{algorithm} \baselineskip 14pt
\vspace{-6pt}
\begin{Lemma} \label{lemma:determine_min_for_attach_g}
	$\hat{min}(v)=\hat{min}(v_l)+\hat{min}(v_r)$.
\end{Lemma}
\begin{proof}
	According to Lemma~\ref{lemma:attach >= true_for_min}, we have $\hat{min}(v) \ge \hat{min}(v_l)+\hat{min}(v_r)$. Therefore, it remains to show that $\hat{min}(v) \le \hat{min}(v_l)+\hat{min}(v_r)$. Let $D_l\in\hat D_{\hat\beta(v_l)}(v_l)$, $D_r \in\hat D_{\hat\alpha(v_r)}(v_r)$ and $D = D_l \cap D_r$. Notice that we have $\hat\beta(v_l) > 0$, for otherwise either condition $D_1$ or condition $D_2$ is true, which is a contradiction to our assumption. Then, one can verify that $\hat{V}(v)-\hat{TS}(v)\subseteq N_{\hat{G}(v)}[D]$ and there exists a vertex set $X \subseteq D \cap \hat{TS}(v)$ such that $G[D - X]$ has a perfect matching with $|X| = \hat\beta(v_l)-\hat\alpha(v_r)$. It follows that $\hat{min}(v)\leq|D| = \hat{min}(v_l) + \hat{min}(v_r)$.
\end{proof}

The following auxiliary lemma is needed in the proof of Lemmas~\ref{lemma:determine_alpha_for_attach_g} and \ref{lemma:determine_beta_for_attach_g}. 
\smallskip
\begin{Lemma} \label{corollary:attach_g min is true min}
	If $D\in \Psi(v)$, then we have $D\in \Psi(\ddot{v})$.
\end{Lemma}
\begin{proof} 
	Suppose that $D \in \Psi(v) \cap \hat D_k(v)$ and $X$ is a set of corresponding unpaired vertices with $|X| = k$. Sine $\hat{TS}(v)\subseteq\hat{TS}(\ddot{v})$, we have $\hat{V}(\ddot{v}) - \hat{TS}(\ddot{v})\subseteq \hat{V}(v) - \hat{TS}(v)$.  Therefore, one can see that $\hat{V}(\ddot{v})-\hat{TS}(\ddot{v})\subseteq N_{\hat{G}(\ddot{v})}[D]$ and  $G[D - X]$ contains a perfect matching with $|X| = k$. Moreover, because of $\hat{min}(v) = \hat{min}(\ddot{v})$ by Lemmas~\ref{lemma:determine_min_for_true} and \ref{lemma:determine_min_for_attach_g}, we have $D\in\Psi(\ddot{v})$.
\end{proof}

\begin{Lemma} \label{lemma:determine_alpha_for_attach_g}
	$
	\hat\alpha(v) =
	\begin{cases}
		\hat \alpha(v_l)- \hat \beta(v_r) & \text{if $\hat \alpha(v_l)> \hat \beta(v_r)$},\\
		|\hat\alpha(v_l)-\hat\alpha(v_r)| \hspace{-5pt } \mod 2   & \text{otherwise.}
	\end{cases}
	$
\end{Lemma}
\begin{proof}
	Let us first consider the case when $\hat \alpha(v_l) > \hat \beta(v_r)$. Suppose that $D \in \hat D_{\hat\alpha(v)}(v)$. Then, by Lemma~\ref{corollary:attach_g min is true min}, we also have $D\in\Psi(\ddot{v})$. This implies that $\hat\alpha(v)\ge\hat\alpha(\ddot{v})$. Combining it further with Lemma~\ref{lemma:determine_alpha_for_true}, one can see that $\hat\alpha(v) \ge\hat\alpha(\ddot{v}) = \hat\alpha(v_l)-\hat\beta(v_r)$. 
	
	Next, we show that $\hat\alpha(v)\le\hat\alpha(v_l)-\hat\beta(v_r)$. Let $D_l \in \hat D_{\hat\alpha}(v_l)$, $D_r\in\hat D_{\hat\beta}(v_r)$, and $D = D_l\cup D_r$. Since $\hat \alpha(v_l) > 0$, one can verify that $\hat{V}(v)-\hat{TS}(v)\subseteq N_{\hat{G}(v)}[D]$ and there exists a vertex set $X\subseteq D \cap\hat{TS}(v)$ such that $G[D - X]$ contains a perfect matching with $|X| = \hat\alpha(v_l) - \hat\beta(v_r)$. Thus, $\hat\alpha(v)\le\hat\alpha(v_l)-\hat\beta(v_r)$. This completes the proof for the case when $\hat \alpha(v_l) > \hat \beta(v_r)$. Using similar arguments, one can show that the equation still holds for the remaining case, hence we omit the details here.
\end{proof}

\begin{Lemma} \label{lemma:determine_beta_for_attach_g}
	$\hat\beta(v) = \hat\beta(v_l)-\hat\alpha(v_r)$.
\end{Lemma}
\begin{proof}
	We first show that $\hat\beta(v) \ge \hat\beta(v_l)-\hat\alpha(v_r)$. Let $D_l \in \hat D_{\hat\beta}(v_l)$, $D_r\in\hat D_{\hat\alpha}(v_r)$, and $D = D_l\cup D_r$. Notice that we have $\hat\beta(v_l) > 0$, for otherwise either condition $D_1$ or condition $D_2$ is true, which is a contradiction to our assumption. Then, one can verify that $\hat{V}(v)-\hat{TS}(v)\subseteq N_{\hat{G}(v)}[D]$ and there exists a vertex set $X\subseteq D \cap\hat{TS}(v)$ such that $G[D - X]$ contains a perfect matching with $|X| = \hat\beta(v_l) - \hat\alpha(v_r)$. Thus, $\hat\beta(v)\geq \hat\beta(v_l)-\hat\alpha(v_r)$.
	
	Then, we show that $\hat\beta(v) \leq \hat\beta(v_l)-\hat\alpha(v_r)$. Suppose that $D \in \hat D_{\hat\beta}(v_l)$, $D_l = D \cap \hat{V}(v_l)$, $D_r = D \cap \hat{V}(v_r)$, and $X \subseteq D \cap \hat{TS}(v)$ is a vertex sets such that $G[D - X]$ has a perfect matching $M$ with size $|X|=\hat\beta(v)$. We also assume that $M$ contains exactly $h$ edges $(x_1, y_1), (x_2, y_2),\ldots, (x_h, y_h)$ such that $x_i \in \hat{TS}(v_l)$ and $y_i \in \hat{TS}(v_r)$ for $i \le h$. Notice that $D\in\Psi(\ddot{v})$ according to Lemma~\ref{corollary:attach_g min is true min}. Applying this result to Lemmas~\ref{lemma:r_and_l_are_minimized_for_true} and~\ref{lemma:true min merge} leads to two sets $D_l\in\Psi(v_l)\cap\hat D_{\hat\beta(v)+h}(v_l)$ and $D_r\in\Psi(v_r)\cap\hat D_{h}(v_r)$ exist. Consequently, we have $\hat\beta(v)+h\le\hat\beta(v_l)$ and $h\ge\hat\alpha(v_r)$. Furthermore, substituting the latter inequality into the former inequality yields $\hat\beta(v) \le \hat\beta(v_l) - h \le \hat\beta(v_l)-\hat\alpha(v_r)$.
\end{proof}

\smallskip

Combining Lemmas~\ref{lemma:determine_min_for_attach_g},~\ref{lemma:determine_alpha_for_attach_g}, and~\ref{lemma:determine_beta_for_attach_g}, we can get the main result of Procedure~\ref{algo:determine_values_for_attach_g}.

\begin{Lemma}\label{lemma:determine_values_for_attach_g}
	If conditions $D_1$ and $D_2$ are false, then Procedure~\ref{algo:determine_values_for_attach_g} determines $\hat{min}(v)$, $\hat\alpha(v)$, and $\hat\beta(v)$ in $O(1)$ time.
\end{Lemma}

\paragraph{Equation~(\ref{eq2}) holds for case $C_1$}  \label{subsection:Equation_holds_for_attach_g}

~\vspace{7pt}

Below, we show Equation~(\ref{eq2}) by the strategy mentioned in Corollary~\ref{corollary:proof_strategy}. First we prove two auxiliary lemmas concerning the properties of $\hat\gamma_0(v)$.

\begin{Lemma} \label{lemma:attach_g gamma_h(G_L) + gamma_h(G_R) = min {gamma_k(G_L) + gamma_k(G_R)}} 
	$\hat\gamma_0(v_l)+\hat\gamma_0(v_r) \ge \min \setof{\hat\gamma_i(v_l)+\hat\gamma_i(v_r) \mid  i \in \setof{1,2} }$.
\end{Lemma}
\begin{proof}
	For the case when $\hat\alpha(v_l) > 0$, we prove the assertion by showing that  $\hat\gamma_0(v_l) + \hat\gamma_0(v_r) \ge \hat\gamma_1(v_l) + \hat\gamma_1(v_r)$. Notice that Equation~(\ref{eq2}) holds for $v_l$ by the induction hypothesis. Therefore, we get $\hat\gamma_0(v_l) = \hat\gamma_1(v_l) + 1$. Further, $\hat\gamma_0(v_r) \ge \hat\gamma_1(v_r) - 1$ by Lemma~\ref{lemma:k=k+-1}. Consequently, $\hat\gamma_0(v_l) + \hat\gamma_0(v_r) \ge \hat\gamma_1(v_l) + \hat\gamma_1(v_r)$. Using similar arguments as above, we can also get $\hat\gamma_0(v_l)+\hat\gamma_0(v_r)\ge\hat\gamma_1(v_l)+\hat\gamma_1(v_r)$ for the case when $\hat\alpha(v_r)>0$. Moreover, since condition $D_2$ is false, it remains to discuss the case when $\hat\alpha(v_l) = 0, \hat\beta(v_l)>0$, $\hat\alpha(v_r) = 0$, and $\hat\beta(v_r) > 0$. Applying Equation~(\ref{eq2}) again yields $\hat\gamma_0(v_l)=\hat\gamma_2(v_l)$ and $\hat\gamma_0(v_r)=\hat\gamma_2(v_r)$. Hence, the assertion of the lemma follows as $\hat\gamma_0(v_l)+\hat\gamma_0(v_r)=\hat\gamma_2(v_l)+\hat\gamma_2(v_r)$. 
\end{proof}

\begin{Lemma} \label{lemma:attach_g gamma_0(G)=min {gamma_k(G_L) + gamma_k(G_R)}} 
	$\hat\gamma_0(v) = \min \setof{\hat\gamma_i(v_l)+\hat\gamma_i(v_r) \mid 0\le i \le min\text{-}ts}$.
\end{Lemma}
\begin{proof}
	We can get $\hat\gamma_0(v) \ge \hat\gamma_0(\ddot{v}) = \min \setof{\hat\gamma_i(v_l)+\hat\gamma_i(v_r) \mid 0\le i \le min\text{-}ts}$ by Lemmas~\ref{lemma:true gamma_0(G)=min {gamma_k(G_L) + gamma_k(G_R)}} and \ref{lemma:attach >= true for gamma_k}. So, it remains to show that $\hat\gamma_0(v) \le \min \{\hat\gamma_i(v_l)+\hat\gamma_i(v_r) \mid 1\le i \le min\text{-}ts\}$ by Lemma~\ref{lemma:attach_g gamma_h(G_L) + gamma_h(G_R) = min {gamma_k(G_L) + gamma_k(G_R)}}. Let $D_l \in \hat D_i(v_l)$, $D_r \in \hat D_i(v_r)$, and $D = D_l\cup D_r$ with $1 \le i\le min\text{-}ts$. Notice that $D_l \cap \hat{TS}(v_l) \not = \emptyset$ as we have $i \ge 1$. Thus, one can verify that $\hat{V}(v)-\hat{TS}(v)\subseteq N_{\hat{G}(v)}[D]$ and $G[D]$ contains a perfect matching. This implies that $\hat\gamma_0(v) \le \min \setof{\hat\gamma_i(v_l)+\hat\gamma_i(v_r) \mid 1\le i \le min\text{-}ts}$. 
\end{proof}

\begin{Lemma} \label{lemma:attach_g gamma_0(G)}
	$\hat\gamma_0(v)=\hat{min}(v)+\hat\alpha(v)$.
\end{Lemma}
\begin{proof}
	Since condition $D_1$ is false, one can verify that $\hat\alpha(\ddot v) = \hat\alpha(v)$ by Lemmas~\ref{lemma:determine_alpha_for_true} and \ref{lemma:determine_alpha_for_attach_g}. Thus, the equation holds, as a consequence of	
	
	\vspace{-20pt}
	\begin{eqnarray*}
		\hspace{-5pt}
		\hat\gamma_0(v) & = & \min \setof{\hat\gamma_k(v_l) + \hat\gamma_k(v_r) \mid 0 \le k \le min\text{-}ts} \text{~~~~(by Lemma~\ref{lemma:attach_g gamma_0(G)=min {gamma_k(G_L) + gamma_k(G_R)}})}\\
		& = & \hat\gamma_{0}(\ddot v) \text{~~~~(by Lemma~\ref{lemma:true gamma_0(G)=min {gamma_k(G_L) + gamma_k(G_R)}})}\\
		& = & \hat{min}(\ddot v)+\hat\alpha(\ddot v)	\text{~~~~(by Lemma~\ref{lemma:true gamma_0(G)})}\\
		& = & \hat{min}(v)+\hat\alpha(\ddot v)	\text{~~~~(by Lemmas~\ref{lemma:determine_min_for_true} and \ref{lemma:determine_min_for_attach_g})}
		\vspace{15pt}\\
		& = & \hat{min}(v)+\hat\alpha(v).	\text{~~~(by Lemmas~\ref{lemma:determine_alpha_for_true} and \ref{lemma:determine_alpha_for_attach_g})}
	\end{eqnarray*}
	
	\vspace{-11pt}
\end{proof}

\vspace{0.5pt}
\begin{Lemma} \label{lemma:attach_g gamma_ts(G)}
	$\hat\gamma_{|\hat{TS}(v)|}(v)=\hat{min}(v)+ |\hat{TS}(v)|-\hat\beta(v)$.
\end{Lemma}
\begin{proof}
	Suppose that $D\in \hat D_{|\hat{TS}(v)|}(v)$, $D_l = D \cap \hat{V}(v_l)$, and $D_r = D \cap \hat{V}(v_r)$. 
	Suppose that $M$ is a matching corresponding to $D$. Then, $M$ contains no edge $(x, y)$ such that $x \in \hat{TS}(v_l)$ and $y \in \hat{TS}(v_r)$. Thus, by Lemma~\ref{lemma:r_and_l_are_minimized_for_attach}, we have $D_l \in \hat D_{|\hat{TS}(v)|}(v_l)$ and $D_r \in \hat D_0(v_r)$. Notice that Equation~(\ref{eq2}) holds fo $v_l$ and $v_r$ according to the induction hypothesis. Then, the claimed result follows, as a consequence of
	
	\vspace{-18pt}
	
	\begin{eqnarray*}
		\hat\gamma_{|\hat{TS}(v)|}(v) 
		& = & \hat\gamma_{|\hat{TS}(v_l)|}(v_l)+\hat\gamma_0(v_r)\text{~~~~(by Lemma~\ref{lemma:r_and_l_are_minimized_for_attach})}\\
		& = & \hat{min}(v_l)+ |\hat{TS}(v_l)|-\hat\beta(v_l) +   
		\hat{min}(v_r)+\hat\alpha(v_r) \text{~~~~(by Equation~(\ref{eq2}))}\\
		& = & \hat{min}(v)+ |\hat{TS}(v)|-\hat\beta(v). 
		\text{~~~~(by Lemmas~\ref{lemma:determine_min_for_attach_g} and~\ref{lemma:determine_beta_for_attach_g}})
	\end{eqnarray*}
	
	\vspace{-6pt}
\end{proof}

\begin{Lemma} \label{lemma:attach_g gamma_alpha+2i(G)=gamma_min(G)}
	$\hat\gamma_{\hat\alpha(v)+2i}(v) = \hat{min}(v)$ for $0 \le i \le (\hat\beta(v)-\hat\alpha(v))/2$.
\end{Lemma}
\begin{proof}
	The proof of this result follows analogous steps as those in the demonstration of Lemma~\ref{lemma:true gamma_alpha+2i(G)=gamma_min(G)}, so we omit it.
\end{proof}

\vspace{8pt}
Combining Corollary \ref{corollary:proof_strategy} and Lemmas~\ref{lemma:attach_g gamma_0(G)}--\ref{lemma:attach_g gamma_alpha+2i(G)=gamma_min(G)}, we obtain the following main result for case $C_1$.

\begin{Lemma} \label{lemma:eq2_holds_for_attach_g}
	If conditions $D_1$ and $D_2$ are false, then Equation~(\ref{eq2}) holds for $v$.
\end{Lemma}

\smallskip
\subsubsection{Case $C_2$: condition $D_1$ is true} 
\label{subsubsection:C_2}
\ 
\vspace{9pt}
\newline
In this subsection, we would like to propose a procedure to determine the values $\hat{min}(v)$, $\hat\alpha(v)$, and $\hat\beta(v)$, and show Equation~(\ref{eq2}) holds.

\paragraph{Determine $\hat{min}(v)$, $\hat\alpha(v)$, and $\hat\beta(v)$ for case $C_2$} \label{subsection:determine_domination-number_for_case_C_2}
\vspace{-16pt}
\floatname{algorithm}{Procedure}
\begin{algorithm}[htb]
	\caption{Determine the values $\hat{min}(v)$, $\hat\alpha(v)$, and $\hat\beta(v)$ for Case $C_2$}
	\label{algo:determine_values_for_attach_s1}
	\begin{algorithmic} [1]
		\baselineskip 14pt
		\REQUIRE $\hat{min}(v_i)$, $\hat\alpha(v_i)$, and $\hat\beta(v_i)$ with $i \in \setof{l,r}$.
		
		\ENSURE $\hat{min}(v)$, $\hat\alpha(v)$, and $\hat\beta(v)$.
		
		\STATE let $\hat{min}(v) \leftarrow \hat{min}(v_l) + \hat{min}(v_r) + \hat\alpha(v_r) - \hat\beta(v_l)$;
		
		\STATE let $\hat\alpha(v) \leftarrow 0$;
		
		\STATE let $\hat\beta(v) \leftarrow 0$;
		
		\RETURN $\hat{min}(v)$, $\hat\alpha(v)$, and $\hat\beta(v)$.
	\end{algorithmic}
\end{algorithm} \baselineskip 14pt

\vspace{-6pt}

\begin{Lemma} \label{lemma:determine_beta_for_attach_s1}
	$\hat\beta(v)= 0$.
\end{Lemma}
\begin{proof}
	Let $D\in\hat D_{\hat\beta(v)}(v)$, $D_l = D \cap \hat{V}(v_l)$ and $D_r = D \cap \hat{V}(v_r)$. Assume that $M$ is a matching corresponding to $D$ and $M$ contains $h$ edges $(x_i, y_i)$ such that $x_i \in \hat{TS}(v_l)$ and $y_i \in \hat{TS}(v_r)$ for $1 \le i \le h$. Suppose to the contrary that $\hat\beta(v)>0$. Then, by Lemma~\ref{lemma:r_and_l_are_minimized_for_attach}, we can get $D_l \in\hat D_{\hat\beta(v)+h}(v_l)$ and $D_r \in\hat D_{h}(v_r)$. To get a contradiction, we first consider the case when $h < \hat\alpha(v_r)$. Notice that $0 < h + 1 \le \hat\alpha(v_r) \le |\hat{TS}(v_r)|$. Thus, there exists a set $D'_r\in\hat D_{h+1}(v_r)$. Moreover, since Equation~(\ref{eq2}) holds for $v_r$ according to the induction hypothesis, we have $|D'_r| = \hat\gamma_h(v_r) - 1$. One can verify that $D' = D_l \cup D'_r$ is a dominating set of $\hat V(v)-\hat{TS}(v)$ and there exists a vertex set $X' \subseteq D' \cap\hat{TS}(v)$ such that $G[D' - X']$ contains a perfect matching with $|X'| = \hat\beta(v) - 1$. These imply that $\hat{min}(v)=|D| > |D'|$, a contradiction.
	
	Next, we consider the case when $h \ge \hat\alpha(v_r)$. Notice that condition $D_1$ is true and so we can get $\hat\beta(v_l) < \hat\alpha(v_r) \le h$. It follows that $\hat\beta(v_l) < \hat\alpha(v_r) \le \hat\beta(v) + h-1 < |\hat{TS}(v_l)|$. Using a similar method of the above arguments, one can show that there exists a vertex set $D'_l\in\hat D_{\hat\beta(v)+h-1}(v_l)$ and $|D'_l|=\hat\gamma_h(v_l)-1$. Meanwhile, one can verify that $D' = D'_l\cup D_r$ is a dominating set of $\hat V(v)-\hat{TS}(v)$ and there exists a vertex set $X'\subseteq D'\cap\hat{TS}(v)$ such that $G[D' - X']$ contains a perfect matching with $|X'| = \hat\beta(v) - 1$. Again, these imply that $\hat{min}(v)=|D| > |D'|$, also a contradiction. So we have that $\hat\beta(v)=0$. 
\end{proof}

\begin{Lemma} \label{lemma:determine_alpha_for_attach_s1}
	$\hat\alpha(v)=0$
\end{Lemma}
\begin{proof}
	Immediate from Lemma~\ref{lemma:determine_beta_for_attach_s1}
\end{proof}

The above lemma implies that $\hat{min}(v) = \hat\gamma_0(v)$. Thus, in order to prove the correctness of $\hat{min}(v)$, we first present two auxiliary lemmas concerning the properties of $\hat\gamma_0(v)$. Moreover, since their proofs are quite similar to  Lemmas~\ref{lemma:attach_g gamma_h(G_L) + gamma_h(G_R) = min {gamma_k(G_L) + gamma_k(G_R)}} and \ref{lemma:attach_g gamma_0(G)=min {gamma_k(G_L) + gamma_k(G_R)}}, respectively, we shall omit them.

\smallskip
\begin{Lemma} \label{lemma:attach_s1 gamma_h(G_L) + gamma_h(G_R) = min {gamma_k(G_L) + gamma_k(G_R)}} 
	$\hat\gamma_0(v_l)+\hat\gamma_0(v_r) \ge \hat\gamma_1(v_l)+\hat\gamma_1(v_r)$.
\end{Lemma}

\smallskip 

\begin{Lemma} \label{lemma:attach_s1 gamma_0(G)=min {gamma_k(G_L) + gamma_k(G_R)}} 
	$\hat\gamma_0(v) = \min \setof{\hat\gamma_i(v_l)+\hat\gamma_i(v_r) \mid 0\le i \le min\text{-}ts}$.
\end{Lemma}

\smallskip
\begin{Lemma} \label{lemma:determine_min_for_attach_s1}
	$\hat{min}(v) = \hat{min}(v_l) + \hat{min}(v_r) + \hat\alpha(v_r) - \hat\beta(v_l)$.
\end{Lemma}
\begin{proof}
	Recall that $\hat{min}(v) = \hat\gamma_{\hat\alpha(v)}(v)$. Combining it further with Lemma~\ref{lemma:determine_alpha_for_attach_s1}, we can get $\hat{min}(v) 
	= \hat\gamma_0(v)$. Then, the equation holds, as a consequence of
	
	\vspace{-18pt}
	
	\begin{eqnarray*}
		\hat{min}(v) 
		& = & \hat\gamma_0(v)\text{~~~~(by Lemma~\ref{lemma:determine_alpha_for_attach_s1})}\\
		& = & \min \{\hat\gamma_i(v_l)+\hat\gamma_i(v_r) \mid 0\le i \le min\text{-}ts\} \text{~~~~(by Lemmas~\ref{lemma:attach_s1 gamma_0(G)=min {gamma_k(G_L) + gamma_k(G_R)}})}\\
		& = & \hat\gamma_0(\ddot{v})\text{~~~~(by Lemma~\ref{lemma:attach >= true_for_min})}  \\ 
		& = & \hat{min}(\ddot{v})+ \hat\alpha(\ddot{v})\text{~~~~(by Lemma~\ref{lemma:true gamma_0(G)})}\\
		& = & \hat{min}(v_l) + \hat{min}(v_r) + \hat\alpha(v_r) - \hat\beta(v_l).\text{~~~~(by Lemmas~\ref{lemma:determine_min_for_true} and~\ref{lemma:determine_alpha_for_true})}
	\end{eqnarray*}
	
	\vspace{-18pt}
\end{proof}

\smallskip
Combining above discussion, we have the main result of Procedure~\ref{algo:determine_values_for_attach_s1}.

\begin{Lemma}\label{lemma:determine_values_for_attach_s1}
	If condition $D_1$ is true, then Procedure~\ref{algo:determine_values_for_attach_s1} determines $\hat{min}(v)$, $\hat\alpha(v)$, and $\hat\beta(v)$ in $O(1)$ time.	
\end{Lemma}

\vspace{-9pt}
\paragraph{Equation~(\ref{eq2}) holds for case $C_2$} \label{subsection:Equation_holds_for_attach_s1}

~\vspace{7pt}

In the following, we show that Equation~(\ref{eq2}) holds according to the strategy mentioned in Corollary~\ref{corollary:proof_strategy}.

\begin{Lemma} \label{corollary:attach_s1 gamma_0(G)}
	$\hat\gamma_0(v)=\hat{min}(v)+\hat\alpha(v)$.
\end{Lemma}
\begin{proof}
	By the definition of $\hat\alpha(v)$, $\hat{min}(v) = \hat\gamma_{\hat\alpha(v)}(v)$. Moreover, since  $\hat\alpha(v) = 0$ according to Lemma~\ref{lemma:determine_alpha_for_attach_s1}, we can get $\hat\gamma_0(v)=\hat{min}(v)+\hat\alpha(v)$.
\end{proof}

\begin{Lemma} \label{lemma:attach_s1 gamma_ts(G)}
	$\hat\gamma_{|\hat{TS}(v)|}(v)=\hat{min}(v)+ |\hat{TS}(v)|-\hat\beta(v)$.
\end{Lemma}
\begin{proof}
	Let $D\in \hat D_{|\hat{TS}(v)|}(v)$, $D_l = D \cap \hat{V}(v_l)$, and $D_r = D \cap \hat{V}(v_r)$. 
	Suppose that $M$ is a matching corresponding to $D$. Then, $M$ contains no edge $(x, y)$ such that $x \in \hat{TS}(v_l)$ and $y \in \hat{TS}(v_r)$.
	Thus, by Lemma~\ref{lemma:r_and_l_are_minimized_for_attach}, we have $D_l \in \hat D_{|\hat{TS}(v)|}(v_l)$ and $D_r \in \hat D_0(v_r)$. Moreover, since Equation~(\ref{eq2}) holds for $v_l$ and $v_r$ by the inductive hypothesis, the statement now follows, as a consequence of
	
	\vspace{-18pt}
	
	\begin{eqnarray*}
		\hat\gamma_{|\hat{TS}(v)|}(v) 
		& = & \hat\gamma_{|\hat{TS}(v_l)|}(v_l)+\hat\gamma_0(v_r)\text{~~~~(by Lemma~\ref{lemma:r_and_l_are_minimized_for_attach})}\\
		& = & \hat{min}(v_l)+ |\hat{TS}(v_l)|-\hat\beta(v_l) +   
		\hat{min}(v_r)+\hat\alpha(v_r) \text{~~~~(by Equation~(\ref{eq2}))}\\
		& = & \hat{min}(v)+ |\hat{TS}(v)|-\hat\beta(v). 
		\text{~~~~(by Lemmas~\ref{lemma:determine_beta_for_attach_s1} and~\ref{lemma:determine_min_for_attach_s1}})
	\end{eqnarray*}
	
	\vspace{-9pt}
\end{proof}

\begin{Lemma} \label{lemma:attach_s1 gamma_alpha+2i(G)=gamma_min(G)}
	$\hat\gamma_{\hat\alpha(v)+2i}(v) = \hat{min}(v)$ for $0 \le i \le (\hat\beta(v)-\hat\alpha(v))/2$.
\end{Lemma}
\begin{proof}
	The lemma follows directly from lemmas~\ref{lemma:determine_beta_for_attach_s1} and \ref{lemma:determine_alpha_for_attach_s1}.
\end{proof}

\smallskip
As an immediate consequence of Corollary \ref{corollary:proof_strategy} and Lemmas~\ref{corollary:attach_s1 gamma_0(G)}--\ref{lemma:attach_s1 gamma_alpha+2i(G)=gamma_min(G)}, we obtain the main result of this subsection.

\begin{Lemma} \label{lemma:eq2_holds_for_attach_s1}
	If condition $D_1$ is true, then Equation~(\ref{eq2}) holds for $v$.
\end{Lemma}


\subsubsection{Case $C_3$: condition $D_2$ is true} 
\label{subsubsection:C_3}
\ 
\vspace{9pt}
\newline
The aim of this subsection is to propose a procedure to determine the values $\hat{min}(v)$, $\hat\alpha(v)$, and $\hat\beta(v)$, and show Equation~(\ref{eq2}) for Case $C_3$.
To complete this aim we first define that the boolean variable $\empts(v)$ is assigned true if and only if $D\cap\hat{TS}(v)= \emptyset$ for all sets $D$ in $\hat D_0(v)$. Furthermore, the description of the methodology for calculating $\emptp(v)$ can be found in Subsection~\ref{subsection:emptp_empts}.

\floatname{algorithm}{Procedure}
\begin{algorithm}[htb]
	\caption{Determine the values $\hat{min}(v)$, $\hat\alpha(v)$, and $\hat\beta(v)$ for Case $C_3$} 
	\label{algo:determine_values_for_attach_s2}
	\begin{algorithmic} [1]
		\baselineskip 14pt
		\REQUIRE $\hat{min}(v_i)$, $\hat\alpha(v_i)$, and $\hat\beta(v_i)$ with $i \in \setof{l,r}$.
		
		\ENSURE $\hat{min}(v)$, $\hat\alpha(v)$, and $\hat\beta(v)$.
		\IF{$\hat\alpha(v_r)=\hat\beta(v_l)=0$}
		\STATE let $\hat{min}(v) \leftarrow \hat{min}(v_l) + \hat{min}(v_l) + (\empts(v_l)\wedge \emptp(v_r))$;
		
		\STATE let $\hat\alpha(v) \leftarrow \empts(v_l)\wedge\emptp(v_r)$;
		
		\STATE let $\hat\beta(v) \leftarrow \empts(v_l)\wedge\emptp(v_r)$;
		\ELSE
		\STATE let $\hat{min}(v) \leftarrow \hat{min}(v_l) + \hat{min}(v_l)$;
		
		\STATE let $\hat\alpha(v) \leftarrow 0$;
		
		\STATE let $\hat\beta(v) \leftarrow \hat\beta(v_l)$;
		\ENDIF
		\RETURN $\hat{min}(v)$, $\hat\alpha(v)$, and $\hat\beta(v)$.
	\end{algorithmic}
\end{algorithm} \baselineskip 14pt

Below, we show the correctness of Procedure~\ref{algo:determine_values_for_attach_s2}. First, a lemma concerning the value of $(\empts(v_l)\wedge\emptp(v_r))$ is provided.

\begin{Lemma} \label{lemma:empts(v_l)_and_empty(v_r)_for_attach_s2}
	Suppose that $D_l \in \Psi(v_l) \cap\hat D_0(v_l)$ and $D_r \in \Psi(v_r) \cap \hat D_0(v_r)$. The value of $(\empts(v_l)\wedge\emptp(v_r))$ is true if and only if there exists no set $D = D_l \cup D_r$ such that $D$ is a dominating set of $\hat V(v)-\hat{TS}(v)$.
\end{Lemma}
\begin{proof}
	First we show the sufficient condition. If $(\empts(v_l)\wedge\emptp(v_r))$ is true, then one can verify that $\hat{TS}(v_r) \not \subseteq N_{\hat G(v)}[D]$ for all sets $D = D_l \cup D_r$. Consequently, $D$ can not be a dominating set of $\hat V(v)-\hat{TS}(v)$. Next, let us go to the necessity. If $\empts(v_l)=0$, then there exists a vertex set $D_l \in \hat D_0(v_l)$ such that $D_l\cap\hat{TS}(v_l)\neq\emptyset$. It follows that we can construct a vertex set $D = D_l \cup D_r$ such that $\hat V(v)-\hat{TS}(v)\subseteq N_{\hat G(v)}[D]$. Similarly, if $\emptp(v_r)=0$, then there exists a vertex set $D_r \in \hat D_0(v_r)$ such that $\hat{TS}(v_r) \subseteq N_{\hat G(v)}[D_r]$. Thus, a desired vertex set $D = D_l \cup D_r$ such that $\hat V(v)-\hat{TS}(v)\subseteq N_{\hat G(v)}[D]$ can also be built.  
\end{proof}

\begin{Lemma} \label{lemma:determine_min_for_attach_s2}
	If $\hat\alpha(v_r)=\hat\beta(v_l)=0$, then we have $\hat{min}(v)=\hat{min}(v_l)+\hat{min}(v_r)+ (\empts(v_l)\wedge\emptp(v_r))$.
\end{Lemma}
\begin{proof}
	First, we consider the case when $\empts(v_l)\wedge\emptp(v_r)=0$. Clearly, it suffices to show that $\hat{min}(v)=\hat{min}(v_l)+\hat{min}(v_r)$. By Lemma~\ref{lemma:empts(v_l)_and_empty(v_r)_for_attach_s2}, there exist two vertex sets $D_l\in\Psi(v_l)\cap\hat D_0(v_l)$ and $D_r\in\Psi(v_r)\cap\hat D_0(v_r)$ such that $D=D_l\cup D_r$ is a dominating set of $\hat V(v)-\hat{TS}(v)$. Thus, we can get $\hat{min}(v)\le|D|=\hat{min}(v_l)+\hat{min}(v_r)$. Combining it further with Lemma~\ref{lemma:attach >= true_for_min}, we have $\hat{min}(v) = \hat{min}(v_l)+\hat{min}(v_r)$.
	
	Next, we consider the case when $\emptp(v_l)\wedge\emptp(v_r)=1$. Let $D_l\in\hat D_0(v_l)$, $D_r\in\hat D_0(v_r)$, and $v\in\hat{TS}(v)$. Then, one can verify $D=D_l\cup D_r\cup\{v\}$ is a dominating set of $\hat V(v)-\hat{TS}(v)$ and so $\hat{min}(v)\le|D|=\hat{min}(v_l)+\hat{min}(v_r)+1$. Again, combining it further with Lemma~\ref{lemma:attach >= true_for_min}, it remains to show that $\hat{min}(v)\neq\hat{min}(v_l)+\hat{min}(v_r)$. For the purpose of contradiction, we assume that $\hat{min}(v)=\hat{min}(v_l)+\hat{min}(v_r)$. Let $D\in\Psi(v)$, $D_l=D\cap \hat{V}(v_l)$, and $D_r=D\cap \hat{V}(v_r)$. By Lemma~\ref{lemma:determine_min_for_true}, we have  $\hat{min}(v)=\hat{min}(\ddot{v})$. Since $\hat{TS}(v)\subseteq\hat{TS}(\ddot{v})$, we obtain $D\in\Psi{(\ddot{v})}$. Applying it to Lemma~\ref{lemma:true min merge}, we obtain that $D_l\in\Psi(v_l) \cap \hat D_0(v_l)$ as $\hat\beta(v_l)=0$.  Together with Lemma~\ref{lemma:r_and_l_are_minimized_for_true}, it entails that $D_r\in\hat D_0(v_r)\cap\Psi(v_r)$ as $\hat\alpha(v_r)=0$. This contradicts to Lemma~\ref{lemma:empts(v_l)_and_empty(v_r)_for_attach_s2}, as $D$ is a dominating set of $\hat V(v)-\hat{TS}(v)$ when $\emptp(v_l)\wedge\emptp(v_r)=1$. 
\end{proof}

\begin{Lemma} \label{lemma:determine_beta_for_attach_s2}
	If $\hat\alpha(v_r)=\hat\beta(v_l)=0$, then $\hat\beta(v)= \empts(v_l)\wedge\emptp(v_r)$.
\end{Lemma}
\begin{proof}
	We first show that $\hat\beta(v)=0$ when $\empts(v_l)\wedge\emptp(v_r)=0$. Conversely, we assume that $\hat\beta(v)>0$. Let $D\in\hat D_{\hat\beta(v)}(v)$, $D_l=D \cap \hat V((v_l)$, and $D_r = D\cap \hat V(v_r)$. Applying Lemma~\ref{lemma:r_and_l_are_minimized_for_attach}, we get $D_l\in\hat D_{\hat\beta(v)+h}(v_l)$ and $D_r\in\hat D_h(v_r)$ for $h\ge 0$. Since $\hat\beta(v_l)=0$ and Equation~(\ref{eq2}) holds for $v_l$ by the inductive hypothesis, we have $|D_l|=\gamma_{\hat\beta(v)+h}(v_l) > \gamma_0(v_l) = \hat{min}(v_l)$. Hence,  $\hat{min}(v) = |D| = |D_l| + |D_r| > \hat{min}(v_l) + \hat{min}(v_r)$, a contradiction to the result of Lemma~\ref{lemma:determine_min_for_attach_s2}. Thus, we have $\hat\beta(v)=0$.
	
	Next, we show that $\hat\beta(v)=1$ when $\empts(v_l)\wedge\emptp(v_r)=1$. Let $D_l\in\hat D_1(v_l)$, $D_r\in\hat D_0(v_r)$, and $D = D_l \cup D_r$. Notice that Equation~(\ref{eq2}) holds for $v_l$ and $v_r$ by inductive hypothesis. Combining it further with Lemma~\ref{lemma:determine_min_for_attach_s2}, we have $|D| = \hat{min}(v_l) + \hat{min}(v_r) + 1 = \hat{min}(v)$. If follows that $D\in\Psi(v)\cap\hat D_1(v)$ and so $\hat{\beta}(v)\ge 1$. Now, it suffices to show that $\hat{\beta}(v)\le 1$.
	Again, we assume for a contradiction that $\hat\beta(v) > 1$. Let $D\in\hat D_{\hat\beta(v)}(v)$, $D_l=D \cap \hat V(v_l)$, and $D_r = D\cap \hat V(v_r)$. Similarly, applying Lemma~\ref{lemma:r_and_l_are_minimized_for_attach}, we get $D_l\in\hat D_{\hat\beta(v)+h}(v_l)$ and $D_r\in\hat D_h(v_r)$ for $h\ge 0$. Since $\hat\beta(v_l)=0$ and Equation~(\ref{eq2}) holds for $v_l$ by the inductive hypothesis, we have $|D_l|=\gamma_{\hat\beta(v)+h}(v_l)>\gamma_1(v_l)=\hat{min}(v_l)+1$. Thus, $\hat{min}(v) = |D| = |D_l| + |D_r| > \hat{min}(v_l) + \hat{min}(v_r)$ + 1, a contradiction to Lemma~\ref{lemma:determine_min_for_attach_s2}. Hence, $\hat\beta(v)= 1$. 
\end{proof}

\begin{Lemma} \label{lemma:determine_alpha_for_attach_s2}
	If $\hat\alpha(v_r)=\hat\beta(v_l)=0$, then $\hat\alpha(v)=\empts(v_l)\wedge\emptp(v_r)$.
\end{Lemma}
\begin{proof}
	For the case when $\empts(v_l)\wedge\emptp(v_r)=0$, the assertion holds immediately from Lemma~\ref{lemma:determine_beta_for_attach_s2}. For the case when $\empts(v_l)\wedge\emptp(v_r)=1$, we can get $\hat\beta(v)=1$ by Lemma~\ref{lemma:determine_beta_for_attach_s2}. Notice that $\hat\gamma_0(v)\ge\hat\gamma_{\hat\beta(v)}(v)$. Combining it further with Lemma~\ref{lemma:k=k+-1}, we obtain $\hat\gamma_0(v) = \hat\gamma_1(v) + 1 > \hat{min}(v)$. Consequently, $\hat\alpha(v)=1$ and the lemma follows.	
\end{proof}

Next, we show that the procedure correctly computes the values $\hat{min}(v)$, $\hat\alpha(v)$, and $\hat\beta(v)$ for the case when $\hat\alpha(v_l)=\hat\beta(v_r)=0$. Notice that the subcase when $\hat\beta(v_r) = \hat\beta(v_l) = 0$ is handled by the if statement of Procedure~\ref{algo:determine_values_for_attach_s2}. Therefore, without loss of generality, we assume that $\hat\beta(v_l) > 0$ below.

\begin{Lemma} \label{lemma:determine_min_for_attach_s2-2}
	If $\hat\alpha(v_l)=\hat\beta(v_r)=0$, then $\hat{min}(v)=\hat{min}(v_l)+\hat{min}(v_r)$.
\end{Lemma}
\begin{proof}
	According to Lemma~\ref{lemma:attach >= true_for_min}, we can obtain $\hat{min}(v)\ge\hat{min}(v_l)+\hat{min}(v_r)$. Hence, it remains to show that $\hat{min}(v)\le\hat{min}(v_l)+\hat{min}(v_r)$. Let $D_l\in\hat D_{\hat\beta(v_l)}(v_l)$, $D_r\in\hat D_0(v_r)$, and $D=D_l\cup D_r$. Since $\hat\beta(v_l)>0$, one can verify that $\hat V(v)-\hat{TS}(v)\subseteq N_{\hat G(v)}[D]$ and there exists a vertex set $X\subseteq D\cap\hat{TS}(v)$ such that $G[D-X]$ contains a perfect matching with $|X|=\hat\beta(v_l)$. Therefore, $\hat{min}(v)\le|D|=\hat{min}(v_l)+\hat{min}(v_r)$.
\end{proof}

\begin{Lemma} \label{lemma:determine_beta_for_attach_s2-2}
	If $\hat\alpha(v_l)=\hat\beta(v_r)=0$, then $\hat\beta(v)=\hat\beta(v_l)$.
\end{Lemma}
\begin{proof}
	Let $D_l\in\hat D_{\hat\beta(v_l)}(v_l)$, $D_r\in\hat D_0(v_r)$, and $D=D_l\cup D_r$.  Since $\hat\beta(v_l)>0$, one can verify that $\hat V(v)-\hat{TS}(v)\subseteq N_{\hat G(v)}[D]$ and there exists a vertex set $X\subseteq D\cap\hat{TS}(v)$ such that $G[D-X]$ contains a perfect matching with $|X|=\hat\beta(v_l)$. According to Lemma~\ref{lemma:determine_min_for_attach_s2-2}, $|D|=\hat{min}(v_l)+\hat{min}(v_r)=\hat{min}(v)$. Therefore, we can get $D\in\Psi(v)$ and so $\hat\beta(v)\ge\hat\beta(v_l)$. Now, it remains to show that $\hat\beta(v)\le\hat\beta(v_l)$. Suppose that $D\in\hat D_{\hat\beta(v)}(v)$. Then, by Lemmas~\ref{lemma:determine_min_for_true} and \ref{lemma:determine_min_for_attach_s2-2}, we have $|D|=\hat{min}(v_l)+\hat{min}(v_r)=\hat{min}(\ddot{v})$. Moreover, since $\hat{TS}(v)\subseteq\hat{TS}(\ddot{v})$, one can verify that $D\in\Psi(\ddot{v})$. Thus, $\hat\beta(v)\le\hat\beta(\ddot{v})$. Combining it further with Lemma~\ref{lemma:determine_beta_for_true}, we have $\hat\beta(v)\le\hat\beta(\ddot{v}) = \hat\beta(v_l)+\hat\beta(v_r) = \hat\beta(v_l)$.
\end{proof}

To prove $\hat\alpha(v) = 0$, we first show that if there exists a vertex set $D_l\in\hat D_0(v_l)$ such that $D_l\cap\hat{TS}(v_l)\neq\emptyset$, then $\hat\alpha(v) = 0$. After that we will show the existence of $D_l$ in Lemma~\ref{lemma:beta>0->empts=0}.

\begin{Lemma} \label{lemma:determine_alpha_auxiliary_s2-2}
	Let $D_l \in\hat D_0(v_l)$, $D_r \in \hat D_0(v_r)$, and $D=D_l\cup D_r$. If $D_l\cap\hat{TS}(v_l)\neq\emptyset$, then $D\in\Psi(v)\cap\hat D_0(v)$. 
\end{Lemma}
\begin{proof}
	By Lemma~\ref{lemma:determine_min_for_attach_s2-2}, we have $|D| = \hat{min}(v_l) + \hat{min}(v_r) = \hat{min}(v)$. Further, one can verify that $\hat V(v) - \hat{TS}(v) \subseteq N_{\hat G(v)}[D]$ and $G[D]$ contains a perfect matching as $D_l \in\hat D_0(v_l)$ and $D_r \in \hat D_0(v_r)$. Consequently, $D\in \Psi(v)\cap\hat D_0(v)$.
\end{proof}

The following two lemmas are provided to prove the existence of $D_l$. For convenience, we let $T_v$ represent the subtree of $T$ rooted at $v$. 

\begin{Lemma} \label{lemma:alpha>0->empts=0}
	For each vertex $u$ in $T_v - \setof{v}$, if $\hat\alpha(u)>0$, then there exists $D\in\hat D_0(u)$ such that $D\cap\hat{TS}(u)\neq\emptyset$.
\end{Lemma}
\begin{proof}
	Suppose that $D\in\hat D_{\hat\alpha(u)}(u)$ and $G[D - X]$ contains a perfect matching for some  vertex set $X \subseteq D \cap \hat{TS}(u)$ of size $\hat\alpha(u)$. Notice that $D\in \Psi(u)$. Therefore, for each vertex $x_i \in X$ with $1 \le i \le \hat\alpha(u)$, there exists a vertex $y_i$ in $\hat V(u) - \hat{TS}(u)$ such that $N_{\hat G(u)}(y_i) \cap D = \setof{x_i}$. Let $S = D\cup \setof{y_1,y_2,\ldots, y_{\hat\alpha(u)}}$. Clearly, we have $\hat V(u) - \hat{TS}(u) \subseteq N_{\hat G(u)}[S]$ and $G[S]$ contains a perfect matching. Further, since Equation~(\ref{eq2}) holds for $u$ by the inductive hypothesis, we can get $|S| =  \hat{min} + \hat\alpha(u) = \hat\gamma_0(u)$. This implies that $S \in\hat D_0(u)$ and $S\cap\hat{TS}(u)\neq\emptyset$.
\end{proof}

\vspace{0.1pt}
\begin{Lemma} \label{lemma:beta>0->empts=0}
	If $\hat\beta(v)>0$, then there exists $D\in\hat D_0(v)$ such that $D\cap\hat{TS}(v)\neq\emptyset$.
\end{Lemma}
\begin{proof}
	We prove by induction on the height of $v$ in $T$. If the height of $v$ is $0$, then $v$ must be a leaf and so $\hat\beta(v) = 0$. Hence, the statement clearly holds. Now, we assume that $v$ is an internal vertex and $\hat\beta(v)>0$. First, consider the case when $v$ is labeled by true twin operation $\otimes$. Let $D\in\hat D_0(v)$, $D_l=D \cap \hat V(v_l)$, and $D_r = D\cap \hat V(v_r)$. Applying Lemma~\ref{lemma:r_and_l_are_minimized_for_true}, we can get $D_l\in\hat D_h(v_l)$ and $D_r\in\hat D_h(v_r)$ for some $0 \le h \le \min \setof{|\hat{TS}(v_l)|,|\hat{TS}(v_r)|}$. If $h > 0$, then the assertion clearly holds as $D = D_l \cup D_r$. So we may assume that $h = 0$. Due to Lemma~\ref{lemma:determine_beta_for_true}, $\hat\beta(v) = \hat\beta(v_l) + \hat\beta(v_r) > 0$, which implies either $\hat\beta(v_l) > 0$ or $\hat\beta(v_r) > 0$. Hence, without loss of generality, we may assume that $\hat\beta(v_l)> 0 $. By the inductive hypothesis, there exists a vertex set $S_l \in\hat D_0(v_l)$ such that $S_l \cap\hat{TS}(v_l)\neq\emptyset$. Construct $S = S_l \cup D_r$. Since $|S| = |D|$, one can verify that $S\in\hat D_0(v)$ and $S\cap\hat{TS}(v) \neq \emptyset$. Using similar arguments, one can prove that the statement also holds when $v$ is labeled by false operation~$\odot$. So we omit it.
	
	Next, we consider the case when $v$ is labeled by attachment operation~$\oplus$. As described before, the discussion is divided into three subcases: $C_1, C_2$, and $C_3$. Again, since the proof for subcase $C_1$ is quite similar to the proof for true twin operation~$\otimes$, we omit the
	details for brevity. Furthermore, subcase $C_2$ never occurs as we have $\hat\beta(v) = 0$ by Lemma~\ref{lemma:determine_beta_for_attach_s1}, which contradicts to the assumption $\hat\beta(v) > 0$. Consequently, we need only discuss subcase $C_3$ for attachment operation~$\oplus$. When $\hat\beta(v_r)=\hat\alpha(v_l)=0$, the proof is also quite similar to the proof of true twin operation~$\otimes$, so we omit it again. Now, we focus on the situation when $\hat\beta(v_l)=\hat\alpha(v_r)=0$. Combining the Lemmas~\ref{lemma:determine_beta_for_attach_s2} and \ref{lemma:determine_alpha_for_attach_s2}, we can get $\hat\beta(v) = \empts(v_l)\wedge\emptp(v_r) = \hat\alpha(v)$, which implies that $\hat\alpha(v) > 0$. Applying this result to Lemma~\ref{lemma:alpha>0->empts=0}, we obtain a vertex set $D\in\hat D_0(v)$ such that $D\cap\hat{TS}(v)\neq\emptyset$.
\end{proof}

Combining Lemmas~\ref{lemma:determine_alpha_auxiliary_s2-2} and \ref{lemma:beta>0->empts=0}, we have the following result.

\begin{Lemma}\label{lemma:determine_alpha_for_attach_s2-2}
	If $\hat\alpha(v_l)=\hat\beta(v_r)=0$, then $\hat\alpha(v)=0$.
\end{Lemma}

\smallskip
Hence we have the main result of Procedure~\ref{algo:determine_values_for_attach_s2}. Further, combining it with Lemmas~\ref{lemma:determine_values_for_attach_g} and \ref{lemma:determine_values_for_attach_s1}, we conclude that $\hat{min}(v)$, $\hat\alpha(v)$, and $\hat\beta(v)$ can be computed in $O(1)$ time for attachment operation~$\oplus$.

\begin{Lemma}\label{lemma:determine_values_for_attach}
	If $\hat\alpha(v_r)=\hat\beta(v_l)=0$ or $\hat\alpha(v_l)=\hat\beta(v_r)=0$, then Procedure~\ref{algo:determine_values_for_attach_s2} determines $\hat{min}(v)$, $\hat\alpha(v)$, and $\hat\beta(v)$ in $O(1)$ time.
\end{Lemma}

\begin{Lemma}\label{lemma:determine_values_for_attachment}
	Suppose that $v$ is annotated by attachment operation $\oplus$, then $\hat{min}(v)$, $\hat\alpha(v)$, and $\hat\beta(v)$ in $O(1)$ time can be determined in $O(1)$ time.
\end{Lemma}

\paragraph{Equation~(\ref{eq2}) holds for case $C_3$} \label{subsection:Equation_holds_for_attach_s2}

~\vspace{7pt}

Again, we show that Equation~(\ref{eq2}) holds for case $C_3$ according to the strategy mentioned in Corollary~\ref{corollary:proof_strategy} below.

\smallskip
\begin{Lemma} \label{lemma:attach_s2 gamma_0(G)}
	$\hat\gamma_0(v)=\hat{min}(v)+\hat\alpha(v)$.
\end{Lemma}
\begin{proof}
	Lemmas~\ref{lemma:determine_alpha_for_attach_s2} and \ref{lemma:determine_alpha_for_attach_s2-2} imply that the value of $\hat\alpha(v)$ is either $0$ or $1$. Meanwhile, the assertion clearly holds for $\hat\alpha(v)=0$. For now we assume that $\hat\alpha(v)=1$. According to Lemma~\ref{lemma:k=k+-1}, we can obtain that $|\hat\gamma_0(v) - \hat\gamma_1(v)| = 1$. Further, since $\hat\gamma_1(v) = \hat{min}(v)$, one can verify that $\hat\gamma_0(v) = \hat\gamma_1(v)+1  = \hat{min}(v)+\hat\alpha(v)$. The lemma then follows.
\end{proof}

\begin{Lemma} \label{lemma:attach_s2 gamma_ts(G)}
	$\hat\gamma_{|\hat{TS}(v)|}(v)=\hat{min}(v)+ |\hat{TS}(v)|-\hat\beta(v)$.
\end{Lemma}
\begin{proof}
	By using similar arguments as in Lemma~\ref{lemma:attach_g gamma_ts(G)}, this assertion can be proven.
\end{proof}

\begin{Lemma} \label{lemma:attach_s2 gamma_alpha+2i(G)=gamma_min(G)}
	$\hat\gamma_{\hat\alpha(v)+2i}(v) = \hat{min}(v)$ for $0 \le i \le (\hat\beta(v)-\hat\alpha(v))/2$.
\end{Lemma}
\begin{proof}
	For the case when $\hat\alpha(v_r) = \hat\beta(v_l) = 0$, the lemma follows directly from lemmas~\ref{lemma:determine_beta_for_attach_s2} and \ref{lemma:determine_alpha_for_attach_s2} as $\hat\alpha(v) = \hat\beta(v)$. On the other hand, for the case when $\hat\alpha(v_l) = \hat\beta(v_r) = 0$, the proof follows analogous steps to those of Lemma~\ref{lemma:true gamma_alpha+2i(G)=gamma_min(G)}, so we omit it.
\end{proof}

Now, as a consequence of Corollary \ref{corollary:proof_strategy} and Lemmas~\ref{lemma:attach_s2 gamma_0(G)}--\ref{lemma:attach_s2 gamma_alpha+2i(G)=gamma_min(G)}, we obtain the result of Lemma~\ref{lemma:eq2_holds_for_attach_s2}. Further, combining it with  Lemmas~\ref{lemma:eq2_holds_for_attach_g} and \ref{lemma:eq2_holds_for_attach_s1}, we conclude that Equation~(\ref{eq2}) holds for attachment operation $\oplus$.  


\begin{Lemma} \label{lemma:eq2_holds_for_attach_s2}
	If condition $D_2$ is true, then Equation~(\ref{eq2}) holds for $v$.
\end{Lemma}

\begin{Lemma} \label{lemma:eq2_holds_for_attachment}
	Suppose that $v$ is labeled by attachment operation $\oplus$ with left child $v_l$ and right child $v_r$. If Equation~(\ref{eq2}) holds for $v_l$ and $v_r$, then the equation also holds for $v$.
\end{Lemma}

\smallskip
\section{Determine $\emptp(v)$ and $\empts(v)$} \label{subsection:emptp_empts} 
This section provides procedures to determine $\emptp(v)$ and $\empts(v)$ for each internal vertex $v$ in $T$. Recall that we have initially assigned $\emptp(\ell) = 1$ and $\empts(\ell) = 1$ for each leaf $\ell \in T$. Hence, we can iteratively processes the internal vertices in $T$ in a bottom-up manner. Suppose that the values $\emptp(v_i)$ and $\empts(v_i)$ have been determined with $i \in \setof{l,r}$. Three cases are discussed when $v$ is labeled by true twin operation $\otimes$, false twin operation $\odot$, and attachment operation~$\oplus$, respectively, in Subsections~\ref{subsection:emptp_empts_for_true}--\ref{subsection:emptp_empts_for_attach}. First, an auxiliary lemma is given.

\begin{Lemma} \label{lemma:auxiliary_for_emptp_&_empts}
	We have $D\notin\hat D_p(v)$ or $D\cap\hat{TS}(v)=\emptyset$ for all vertex sets $D\in\hat D_0(v)$ if and only if $\emptp(v)\vee\empts(v)=1$. 
\end{Lemma}
\begin{proof}
	We prove the statement by induction on the height of $v$ in $T$. If the height of $v$ is $0$, then $v$ must be a leaf. Hence, we have $\hat D_0(v)=\{\emptyset\}$ and $\hat D_p(v)=\emptyset$, $\empts(v) = 1$ and $\emptp(v) = 1$, the lemma certainly holds. Then, we assume that $v$ is an internal vertex with left child $v_l$ and right child $v_r$. By the induction hypothesis, Lemma~\ref{lemma:auxiliary_for_emptp_&_empts} holds for $v_l$ and $v_r$. Under these conditions, we will show the statement also holds for the cases when $v$ is labeled by true twin operation $\otimes$, false twin operation $\odot$, and attachment operation $\oplus$, respectively, in Lemmas~\ref{lemma:property1_holds_for_true},~\ref{lemma:property1_holds_for_false}, and~\ref{lemma:property1_holds_for_attach}.  Thus, the lemma follows.
\end{proof}

\subsection{Determine $\emptp(v)$ and $\empts(v)$ for true twin operation $\otimes$} \label{subsection:emptp_empts_for_true} 
Suppose that $v$ is an internal vertex of $T$ with left child $v_l$ and right child $v_r$. Suppose further that Lemma~\ref{lemma:auxiliary_for_emptp_&_empts} holds for $v_i$. Below, we first propose a procedure to determine $\emptp(v)$ and $\empts(v)$. Then, we show that Lemma~\ref{lemma:auxiliary_for_emptp_&_empts} also holds for $v$. 

As we can see in Procedure~\ref{algo:determine_empty_&_empts_for_true}, condition $D_2$ restricts how the values $\emptp(v)$ and $\empts(v)$ are determined. Remember that condition $D_2$ holds when $\hat\alpha(v_r)=\hat\beta(v_l)=0$ or $\hat\alpha(v_l)=\hat\beta(v_r)=0$, which was defined in Subsection~\ref{subsection:attachment}. If condition $D_2$ holds, then $\emptp(v)$ and $\empts(v)$ can be computed from $\emptp(v_i)$, and $\empts(v_i)$ with $i\in\{l,r\}$. Otherwise, both values are equal to~$0$.

\smallskip
\begin{algorithm}[htb]
	\caption{Determine $\emptp(v)$ and $\empts(v)$ for true twin operation $\otimes$}\label{algo:determine_empty_&_empts_for_true}
	\begin{algorithmic} [1]
		\baselineskip 14pt
		\REQUIRE $\hat\alpha(v_i)$, $\hat\beta(v_i)$, $\emptp(v_i)$, and $\empts(v_i)$ with $i\in\{l,r\}$.
		\ENSURE  $\emptp(v)$ and $\empts(v)$.
		
		\IF {condition $D_2$ holds}
		\STATE $\emptp(v)\leftarrow (\emptp(v_l)\vee\emptp(v_r))\wedge(\empts(v_l)\vee\empts(v_r))\wedge~$
		
		\hspace{60pt}$(\emptp(v_l)\vee\empts(v_l))\wedge(\emptp(v_r)\vee\empts(v_r))$;
		\STATE $\empts(v)\leftarrow \empts(v_l)\wedge\empts(v_r)$;
		\ELSE
		\STATE $\emptp(v)\leftarrow 0$;
		\STATE $\empts(v)\leftarrow 0$;
		\ENDIF	
		\RETURN $\emptp(v)$ and  $\empts(v)$.
	\end{algorithmic}
\end{algorithm}

Let $D \subseteq \hat V(v)$, $D_l = D\cap\hat V(v_l)$, and $D_r=D\cap\hat V(v_r)$. Two auxiliary lemmas are introduced in order to prove the correctness of Procedure~\ref{algo:determine_empty_&_empts_for_true}. 

\begin{Lemma} \label{lemma:D0_for_true_D2_holds}
	Suppose that condition $D_2$ holds. Then, $D\in\hat D_0(v)$ if and only if $D_l\in\hat D_0(v_l)$ and $D_r\in\hat D_0(v_r)$.
\end{Lemma}
\begin{proof}
	First we show the sufficient condition. Without loss of generality, we can assume that $D\in\hat D_0(v)$ and $\hat\alpha(v_l) = \hat\beta(v_r)=0$. Applying Lemma~\ref{lemma:r_and_l_are_minimized_for_true}, we get $D_l\in\hat D_k(v_l)$ and $D_r\in\hat D_k(v_r)$. Assume for the purpose of contradiction that $k>0$. Since $\hat\alpha(v_l)=0$, we have $|D_l|\geq \hat\gamma_0(v_l)$. Further, since $\hat\beta(v_r)=0$, the value of $|D_r|=\hat{min}(v_r) + k>\hat\gamma_0(v_r)$ can be calculated by Equation~(\ref{eq2}). 
	Combining it with Lemma~\ref{lemma:true gamma_0(G)=min {gamma_k(G_L) + gamma_k(G_R)}}, $\hat\gamma_0(v)=\min \setof{\hat\gamma_i(v_l)+\hat\gamma_i(v_r) \mid 0\le i \le min\text{-}ts}\le\hat\gamma_0(v_l)+\hat\gamma_0(v_r)< |D_l| + |D_r| = |D|$, a contradiction. Consequently, we have $k=0$, which implies $D_l\in\hat D_0(v_l)$ and $D_r\in\hat D_0(v_r)$.	
	
	Let us go to the necessity. Suppose that $D_l\in\hat D_0(v_l)$ and $D_r\in\hat D_0(v_r)$. Since $\hat\alpha(v_l)=\hat\beta(v_r)=0$, we have $|D_l|=\hat{min}(v_l)$ and $|D_r|=\hat{min}(v_r)$. By Lemma~\ref{lemma:determine_min_for_true}, we have $|D|=\hat{min}(v_l)+\hat{min}(v_r)=\hat{min}(v)$. One can verify that $D = D_l \cup D_r$ is a dominating set of $\hat V(v) - \hat{TS}(v)$ such that $\hat G(v)[D]$ contains a perfect matching. It follows that $D\in\hat D_0(v)$.
\end{proof}

\begin{Lemma} \label{lemma:D0_for_true_D2_not_holds}
	If condition $D_2$  is false, there exists a vertex set $D\in\hat D_0(v)\cap\hat D_p(v)$ such that $D\cap\hat{TS}(v)\neq\emptyset$.
\end{Lemma}
\begin{proof}
	Suppose that $D\in\hat D_0(v)$. Then, by Lemma~\ref{lemma:r_and_l_are_minimized_for_true}, we can get $D_l\in\hat D_k(v_l)$ and $D_r\in\hat D_k(v_r)$. Notice that if $k > 0$, then we have $D\in\hat D_p(v)$ and $D\cap\hat{TS}(v)\neq\emptyset$. Thus, it remains to consider the case when $k=0$. In particular by applying Lemma~\ref{lemma:true gamma_0(G)=min {gamma_k(G_L) + gamma_k(G_R)}}, we obtain $\hat\gamma_0(v)=\min \setof{\hat\gamma_i(v_l)+\hat\gamma_i(v_r) \mid 0\le i \le min\text{-}ts}$. Thus, the proof can be completed by showing that, if condition $D_2$ is false, then $\hat\gamma_0(v_l) + \hat\gamma_0(v_r) \ge \min \setof{\hat\gamma_1(v_l) + \hat\gamma_1(v_r), \hat\gamma_2(v_l) + \hat\gamma_2(v_r)}$. Moreover, one can verify that this can be carried out by using similar arguments to those one used in Lemma~\ref{lemma:attach_g gamma_h(G_L) + gamma_h(G_R) = min {gamma_k(G_L) + gamma_k(G_R)}}. Hence, we omit the details here.
\end{proof}

Let $c_1=\emptp(v_l)\vee\emptp(v_r)$, $c_2=\empts(v_l)\vee\empts(v_r)$, $c_3=\emptp(v_l)\vee\empts(v_l)$, and $c_4=\emptp(v_r)\vee\empts(v_r)$. We establish the correctness of Procedure~\ref{algo:determine_empty_&_empts_for_true} in Lemma~\ref{lemma:determine_emptp_for_true} and Lemma~\ref{lemma:determine_empts_for_true}.

\smallskip

\begin{Lemma} \label{lemma:determine_emptp_for_true}
	$
	\emptp(v) =
	\begin{cases}
		c_1\wedge c_2\wedge c_3\wedge c_4 & \text{if condition $D_2$ holds},\\
		0   & \text{otherwise}.
	\end{cases}
	$
\end{Lemma}
\begin{proof}
	Clearly, by Lemma~\ref{lemma:D0_for_true_D2_not_holds}, if condition $D_2$ is false, then there exists a vertex set $D\in\hat D_0(v)\cap\hat D_p(v)$, which implies $\emptp(v)=0$. Hence, we only need to consider the case when condition $D_2$ holds. Suppose that $D\in\hat D_0(v)$. Then, one can see that $D\in \hat D_p(v)$ if and only if $\hat{TS}(v)\subseteq\hat G(v)[D]$. Notice that $v$ is labeled by true operation $\otimes$. Hence, $D\notin \hat D_p(v)$ if and only if $\hat{TS}(v_l)\not\subseteq\hat G(v)[D_l]\cup \hat G(v)[D_r]$ or $\hat{TS}(v_r)\not\subseteq\hat G(v)[D_l]\cup \hat G(v)[D_r]$. Let $p(i)$ denote the open statement, ``$\hat{TS}(v_l)\not\subseteq\hat G(v)[D_i]$'' and $q(j)$ denote the open statement, ``$\hat{TS}(v_r)\not\subseteq\hat G(v)[D_j]$'' with $i,j \in \setof{l,r}$. Since $D\in\hat D_0(v)$, we have $D_l\in\hat D_0(v_l)$ and $D_r\in\hat D_0(v_r)$ by Lemma~\ref{lemma:D0_for_true_D2_holds}. It follows that 
	
	\vspace*{-14pt}	
	\begin{eqnarray*}
		\emptp(v)
		& \Leftrightarrow & \forall D\in\hat D_0(v)~\hat{TS}(v)\not\subseteq\hat G(v)[D]\\
		& \Leftrightarrow & \forall D_l\in\hat D_0(v_l)~\forall D_r\in\hat D_0(v_r),\\
		&~&(\hat{TS}(v_l)\not\subseteq\hat G(v)[D_l]\cup \hat G(v)[D_r]) \text{ or }\\
		&~&(\hat{TS}(v_r)\not\subseteq\hat G(v)[D_l]\cup \hat G(v)[D_r])\\
		& \Leftrightarrow & \forall D_l\in\hat D_0(v_l)~\forall D_r\in\hat D_0(v_r),\\
		&~&(\hat{TS}(v_l)\not\subseteq\hat G(v)[D_l]\text{ or } \hat{TS}(v_r)\not\subseteq\hat G(v)[D_r]) \text{ and }\\
		&~&(\hat{TS}(v_l)\not\subseteq\hat G(v)[D_r]\text{ or } \hat{TS}(v_r)\not\subseteq\hat G(v)[D_l]) \text{ and }\\
		&~&(\hat{TS}(v_l)\not\subseteq\hat G(v)[D_l]\text{ or } \hat{TS}(v_r)\not\subseteq\hat G(v)[D_l]) \text{ and }\\
		&~&(\hat{TS}(v_l)\not\subseteq\hat G(v)[D_r]\text{ or } \hat{TS}(v_r)\not\subseteq\hat G(v)[D_r])\\
		& \Leftrightarrow & \forall D_l\in\hat D_0(v_l)~\forall D_r\in\hat D_0(v_r),\\
		&~&(p(l) \vee q(r)) \wedge (p(r) \vee q(l))  \wedge (p(l) \vee q(l))  \wedge (p(r) \vee q(r))\\
		& \Leftrightarrow & \forall D_l\in\hat D_0(v_l)~\forall D_r\in\hat D_0(v_r),\\
		&~&c_1  \wedge c_2 \wedge (p(l) \vee q(l))  \wedge (p(r) \vee q(r))\\
		& \Leftrightarrow & c_1\wedge c_2\wedge c_3\wedge c_4
	\end{eqnarray*}
	
	\noindent We observe that $\forall D_l\in\hat D_0(v_l)~p(l)\vee q(l) \Leftrightarrow (\forall D_l\in\hat D_0(v_l)~p(l)) \vee (\forall D_l\in\hat D_0(v_l)~q(l))$; and $\forall D_r\in\hat D_0(v_r)~p(r)\vee q(r) \Leftrightarrow (\forall D_r\in\hat D_0(v_r)~p(r)) \vee (\forall D_r\in\hat D_0(v_r)~q(r))$, as Lemma~\ref{lemma:auxiliary_for_emptp_&_empts} works for $v_l$ and $v_r$. Hence the last logical equivalence above holds.
\end{proof}

\begin{Lemma} \label{lemma:determine_empts_for_true}
	$
	\empts(v) =
	\begin{cases}
		\empts(v_l)\wedge\empts(v_r) & \text{if condition $D_2$ holds},\\
		0   & \text{otherwise.}
	\end{cases}
	$
\end{Lemma}
\begin{proof}
	According to Lemma~\ref{lemma:D0_for_true_D2_not_holds}, if condition $D_2$ is false, then there exists a vertex set $D\in\hat D_0(v)$ such that $D\cap\hat{TS}(v)\neq\emptyset$, which implies $\empts(v)=0$. Thus, it remains to consider the case when condition $D_2$ holds. By Lemma~\ref{lemma:D0_for_true_D2_holds}, we have $D\in\hat D_0(v)$ if and only if $D_l\in\hat D_0(v_l)$ and $D_r\in\hat D_0(v_r)$. Furthermore, one can verify that $D\cap\hat{TS}(v)=\emptyset$ if and only if $D_l\cap\hat{TS}(v_l)=\emptyset$ and $D_r\cap\hat{TS}(v_r)=\emptyset$. These imply $\empts(v) = \empts(v_l)\wedge\empts(v_r)$.  
\end{proof}

Applying Lemmas~\ref{lemma:determine_emptp_for_true} and~\ref{lemma:determine_empts_for_true}, we can get the correctness of Procedure~\ref{algo:determine_empty_&_empts_for_true}. Moreover, the running time of the procedure is $O(1)$ as all the steps can be easily implemented in $O(1)$ time. Thus we have the following result for true twin operation $\otimes$. 

\begin{Lemma}\label{lemma:determine_emptp_empts_for_true}
	Suppose that $v$ is labeled by true twin operation $\otimes$, then Procedure~\ref{algo:determine_empty_&_empts_for_true} determines $\emptp(v)$ and $\empts(v)$ in $O(1)$ time.
\end{Lemma}

\smallskip
Below, in order to complete the proof of Lemma~\ref{lemma:auxiliary_for_emptp_&_empts}, we show the lemma holds for true twin operation $\otimes$.

\begin{Lemma}\label{lemma:property1_holds_for_true}
	Suppose that $v$ is labeled by true twin operation $\otimes$. Then, for all vertex sets $D\in\hat D_0(v)$, we have $D\notin\hat D_p(v)$ or $D\cap\hat{TS}(v)=\emptyset$  if and only if $\emptp(v)\vee\empts(v)=1$. 
\end{Lemma}
\begin{proof}
	Clearly, if $\emptp(v)\vee\empts(v)=1$, then we have $D\notin\hat D_p(v)$ or $D\cap\hat{TS}(v)=\emptyset$ for all vertex sets $D\in\hat D_0(v)$. Thus, it remains to show the sufficient condition. Apparently, the proof holds when $\empts(v)=1$. To complete the proof, it suffices to show that if $\empts(v)=0$, then $\emptp(v)=1$.  
	
	Since $\empts(v)=0$, there exists a vertex set $S\in\hat D_0(v)$ such that $S\cap\hat{TS}(v)\neq\emptyset$. Let $S_l=S\cap\hat V(v_l)$ and $S_r=S\cap\hat V(v_r)$. Without loss of generality, we assume that $S_l\cap\hat{TS}(v_l)\neq\emptyset$. Moreover, as $D\notin\hat D_p(v)$ or $D\cap\hat{TS}(v)=\emptyset$ for all vertex sets $D\in\hat D_0(v)$, we can get condition $D_2$ holds by applying Lemma~\ref{lemma:D0_for_true_D2_not_holds}. Combining it further with Lemma~\ref{lemma:D0_for_true_D2_holds}, one can see that $S_l\in\hat D_0(v_l)$ and $S_r\in\hat D_0(v_r)$. Since $S_l\cap\hat{TS}(v_l)\neq\emptyset$, we have $\empts(v_l)=0$. We note that according to Lemma~\ref{lemma:determine_emptp_for_true}, showing $\emptp(v)=1$ is equivalent to showing that $\emptp(v_l)=1$ and $\empts(v_r)=1$. To achieve this aim, we suppose for the purpose of contradiction that $\emptp(v_l)=0$. Then, since Lemma~\ref{lemma:auxiliary_for_emptp_&_empts} holds for $v_l$ and $\emptp(v_l)\vee\empts(v_l)=0$, there exists a vertex set $S'_l\in\hat D_0(v_l)\cap\hat D_p(v_l)$ such that $S'_l\cap\hat{TS}(v_l)\neq\emptyset$. Let $S'=S'_l\cup S_r$ and we have $S' \in\hat D_0(v)$ by Lemma~\ref{lemma:D0_for_true_D2_holds}. Further, one can verify that $S'\in\hat D_p(v)$ and $S'\cap\hat{TS}(v)\neq\emptyset$. This contradicts to the fact that $D\notin\hat D_p(v)$ or $D\cap\hat{TS}(v)=\emptyset$ for all vertex sets $D\in\hat D_0(v)$. Thus, we have $\emptp(v_l)=1$. Using similar arguments as above, one can also show that $\empts(v_r)=1$.
\end{proof}

\subsection{Determine $\emptp(v)$ and $\empts(v)$ for false twin operation $\odot$} \label{subsection:emptp_empts_for_false} 
Suppose that $v$ is an internal vertex of $T$ with left child $v_l$ and right child $v_r$. We further suppose that Lemma~\ref{lemma:auxiliary_for_emptp_&_empts} holds for left child $v_l$ and right child $v_r$. Below, we first propose a procedure to determine $\emptp(v)$ and $\empts(v)$. Then, we will show that Lemma~\ref{lemma:auxiliary_for_emptp_&_empts} also holds for $v$.

\begin{algorithm}[htb]
	\caption{Determine $\emptp(v)$ and $\empts(v)$ for false twin operation $\odot$}\label{algo:determine_empty_&_empts_for_false}
	\begin{algorithmic} [1]
		\baselineskip 14pt
		\REQUIRE $\emptp(v_i)$ and $\empts(v_i)$ with $i\in\{l,r\}$.
		\ENSURE  $\emptp(v)$ and $\empts(v)$.		
		\STATE $\emptp(v)\leftarrow \emptp(v_l)\vee\emptp(v_r)$;
		\STATE $\empts(v)\leftarrow \empts(v_l)\wedge\empts(v_r)$;
		\RETURN $\emptp(v)$ and  $\empts(v)$.
	\end{algorithmic}
\end{algorithm}

Let $D \subseteq \hat V(v)$, $D_l = D\cap\hat V(v_l)$, and $D_r=D\cap\hat V(v_r)$. We begin with an auxiliary lemma for proving the correctness of Procedure~\ref{algo:determine_empty_&_empts_for_false}.

\begin{Lemma} \label{lemma:D0_for_false}
	$D\in\hat D_0(v)$ if and only if $D_l\in\hat D_0(v_l)$ and $D_r\in\hat D_0(v_r)$.
\end{Lemma}
\begin{proof}
	The sufficient condition holds by Lemma~\ref{lemma:r_and_l_are_minimized_for_false}. Hence, it remains to show the necessity. Let $D_l\in\hat D_0(v_l)$ and $D_r\in\hat D_0(v_r)$. One can verify that $D = D_l\cup D_r$ is a dominating set of $\hat V(v) - \hat{TS}(v)$ such that $\hat G(v)[D]$ contains a perfect matching. Therefore, it suffices to show that $|D|=\hat\gamma_0(v)$. By Equation~(\ref{eq2}), $|D|=\hat\gamma_0(v_l)+\hat\gamma_0(v_r) = \hat{min}(v_l)+\hat\alpha(v_l) + \hat{min}(v_r)+\hat\alpha(v_r)$. Meanwhile, according to Lemmas~\ref{lemma:determine_min_for_false}~and~\ref{lemma:determine_alpha_for_false},  $\hat{min}(v)=\hat{min}(v_l)+\hat{min}(v_r)$ and $\hat\alpha(v)=\hat\alpha(v_l)+\hat\alpha(v_r)$, respectively. Thus, applying Equation~(\ref{eq2}) again, one can see that $|D|=\hat{min}(v)+\hat\alpha(v)=\hat\gamma_0(v)$.
\end{proof}

Now let us establish the correctness of Procedure~\ref{algo:determine_empty_&_empts_for_false} by Lemmas~\ref{lemma:determine_emptp_for_false} and \ref{lemma:determine_empts_for_false}.

\begin{Lemma} \label{lemma:determine_emptp_for_false}
	$\emptp(v)=\emptp(v_l)\vee\emptp(v_r)$.
\end{Lemma}
\begin{proof}
	By Lemma~\ref{lemma:D0_for_false}, we have $D\in\hat D_0(v)$ if and only if $D_l\in\hat D_0(v_l)$ and $D_r\in\hat D_0(v_r)$. Moreover, since $v$ is labeled by false twin operation $\odot$, one can verify that $D\notin\hat D_p(v)$ if and only if $D_l\notin\hat D_p(v_l)$ or $D_r\notin\hat D_p(v_r)$. Hence, $\emptp(v)=\emptp(v_l)\vee\emptp(v_r)$.
\end{proof}

\begin{Lemma} \label{lemma:determine_empts_for_false}
	$\empts(v)=\empts(v_l)\wedge\empts(v_r)$.
\end{Lemma}
\begin{proof}
	According to Lemma~\ref{lemma:D0_for_false}, we have $D\in\hat D_0(v)$ if and only if $D_l\in\hat D_0(v_l)$ and $D_r\in\hat D_0(v_r)$. Therefore, one can verify that $D\cap\hat{TS}(v)=\emptyset$ if and only if $D_l\cap\hat{TS}(v_l) = \emptyset$ and $D_r\cap\hat{TS}(v_r) = \emptyset$. It follows that $\empts(v)=\empts(v_l)\wedge\empts(v_r)$.
\end{proof}

According to Lemmas~\ref{lemma:determine_emptp_for_false} and~\ref{lemma:determine_empts_for_false}, we can get the correctness of Procedure~\ref{algo:determine_empty_&_empts_for_false}. Moreover, since each step of the procedure can be implemented in $O(1)$ time naturally, we have the following result. 

\begin{Lemma}\label{lemma:determine_emptp_empts_for_false}
	Suppose that $v$ is annotated by false twin operation $\odot$, then Procedure~\ref{algo:determine_empty_&_empts_for_false} determines $\emptp(v)$ and $\empts(v)$ in $O(1)$ time.
\end{Lemma}

\smallskip
Below, in order to complete the proof of Lemma~\ref{lemma:auxiliary_for_emptp_&_empts}, we show the lemma holds for false twin operation $\odot$.

\begin{Lemma}\label{lemma:property1_holds_for_false}
	Suppose that $v$ is labeled by false twin operation $\odot$. Then, we have $D\notin\hat D_p(v)$ or $D\cap\hat{TS}(v)=\emptyset$ for all vertex sets $D\in\hat D_0(v)$ if and only if $\emptp(v)\vee\empts(v)=1$. 
\end{Lemma}
\begin{proof}
	If $\emptp(v)\vee\empts(v)=1$, then we have $D\notin\hat D_p(v)$ or $D\cap\hat{TS}(v)=\emptyset$ for all vertex sets $D\in\hat D_0(v)$. Therefore, it remains to show the sufficient condition. Further, since the proof holds when $\empts(v)=1$, it suffices to show that if $\empts(v)=0$, then $\emptp(v)=1$.  
	
	Since $\empts(v)=0$, we have $\empts(v_l)=0$ or $\empts(v_r)=0$ by Lemma~\ref{lemma:determine_empts_for_false}. Without loss of generality, we assume that $\empts(v_l)=0$. Suppose for the purpose of contradiction that $\emptp(v)=0$. Then, by Lemma~\ref{lemma:determine_emptp_for_false}, we have $\emptp(v_l)=0$ and $\emptp(v_r)=0$, which implies that there exists a vertex set $S_r\in\hat D_0(v_r)\cap\hat D_p(v_r)$. Moreover, since Lemma~\ref{lemma:auxiliary_for_emptp_&_empts} holds for $v_l$ and $\emptp(v_l)\vee\empts(v_l)=0$, there exists a vertex set $S_l\in\hat D_0(v_l)\cap\hat D_p(v_l)$ such that $S_l\cap\hat{TS}(v_l)\neq\emptyset$. Let $S=S_l\cup S_r$. Then, we have $S \in\hat D_0(v)$ by Lemma~\ref{lemma:D0_for_false}. Further, one can verify that $S\in\hat D_p(v)$ and $S\cap\hat{TS}(v)\neq\emptyset$. This contradicts to the fact that $D\notin\hat D_p(v)$ or $D\cap\hat{TS}(v)=\emptyset$ for all vertex sets $D\in\hat D_0(v)$. Hence, we have $\emptp(v)=1$.
\end{proof}

\subsection{Determine $\emptp(v)$ and $\empts(v)$ for attachment operation $\oplus$} \label{subsection:emptp_empts_for_attach} 
Suppose that Lemma~\ref{lemma:auxiliary_for_emptp_&_empts} holds for left child $v_l$ and right child $v_r$. In this subsection, we first propose a procedure to determine $\emptp(v)$ and $\empts(v)$. And, we will show that Lemma~\ref{lemma:auxiliary_for_emptp_&_empts} also holds for $v$. The procedure determines the values $\emptp(v)$ and $\empts(v)$ depending on condition $D_3$, which is defined as follows:
\vspace{10pt}
\\
\mbox{} \hspace{55pt} $D_3$ : condition $D_2$ holds and $\emptp(v_l)\wedge\empts(v_r) = 0$.  
\vspace{10pt}
\\
\noindent As one can see in Procedure~\ref{algo:determine_empty_&_empts_for_attach},  the determinations of $\emptp(v)$ and $\empts(v)$ are similar to those employed by true twin operation~$\otimes$. The only difference is the value assigned to $\empts(v_r)$ when condition $D_3$ holds. For this reason, properties of true twin operation~$\otimes$ are particularly useful to prove the correctness of Procedure~\ref{algo:determine_empty_&_empts_for_attach}. 

Let $\ddot{v}$ be an internal vertex labeled by true twin operation~$\otimes$ with left child $v_l$ and right child $v_r$. We first introduce one lemma concerning true twin operation~$\otimes$.

\vspace*{-2pt}
\begin{algorithm}[htb]
	\caption{Determine $\emptp(v)$ and $\empts(v)$ for attachment operation $\oplus$}\label{algo:determine_empty_&_empts_for_attach}
	\begin{algorithmic} [1]
		\baselineskip 14pt
		\REQUIRE $\hat\alpha(v_i)$, $\hat\beta(v_i)$, $\emptp(v_i)$, and $\empts(v_i)$ with $i\in\{l,r\}$.
		\ENSURE  $\emptp(v)$ and $\empts(v)$.		
		\IF {condition $D_3$ holds}
		\STATE $\emptp(v)\leftarrow (\emptp(v_l)\vee\emptp(v_r))\wedge(\empts(v_l)\vee\empts(v_r))\wedge~$
		
		\hspace{60pt}$(\emptp(v_l)\vee\empts(v_l))\wedge(\emptp(v_r)\vee\empts(v_r))$;
		\STATE $\empts(v)\leftarrow \empts(v_l)$;
		\ELSE
		\STATE $\empts(v)\leftarrow 0$;
		\STATE $\emptp(v)\leftarrow 0$;
		\ENDIF
		\RETURN $\emptp(v)$ and  $\empts(v)$.
	\end{algorithmic}
\end{algorithm}
\vspace*{-12pt}

\begin{Lemma} \label{lemma:D0_for_attach_D3_holds}
	If condition $D_3$ holds, then $\hat\gamma_0(v)=\hat\gamma_0(\ddot v)$. 
\end{Lemma}
\begin{proof}
	Since condition $D_2$ holds, we have either $\hat\alpha(v_r)=\hat\beta(v_l)=0$ or $\hat\alpha(v_l)=\hat\beta(v_r)=0$. Hence we discuss the two cases separately. First we consider the case when $\hat\alpha(v_r)=\hat\beta(v_l)=0$. Since condition $D_3$ holds, we have $\empts(v_l)\wedge\emptp(v_r)=0$. Further, applying Lemma~\ref{lemma:determine_alpha_for_attach_s2}, we can get $\hat\alpha(v)=0$. Hence,
	
	\vspace*{-15pt}
	\begin{eqnarray*}
		\hspace{-5pt}
		\hat\gamma_0(v) & = & \hat{min}(v)+\hat\alpha(v) 
		\text{~~~~(by Lemma~\ref{lemma:attach_s2 gamma_0(G)})}\\
		& = & \hat{min}(v_l)+\hat{min}(v_r)
		\text{~~~~(by Lemmas~\ref{lemma:determine_min_for_attach_s2} and~\ref{lemma:determine_alpha_for_attach_s2}})\\
		& = & \hat{min}(\ddot v) \text{~~~~(by Lemmas~\ref{lemma:determine_min_for_true})}\\
		& = & \hat{min}(\ddot v)+\hat\alpha(\ddot v) 
		\text{~~~~(by Lemmas~\ref{lemma:determine_alpha_for_true})}\\
		& = & \hat\gamma_0(\ddot v). 
		\text{~~~~(by Lemmas~\ref{lemma:true gamma_0(G)})} 
	\end{eqnarray*} 
	\vspace{-15pt}
	
	\noindent Next, we consider the case when $\hat\alpha(v_l)=\hat\beta(v_r)=0$.	
	Combining the results of Lemmas~\ref{lemma:determine_min_for_attach_s2-2}, \ref{lemma:determine_alpha_for_attach_s2-2}, and \ref{lemma:attach_s2 gamma_0(G)}, we can also obtain $\hat\gamma_0(v)=\hat{min}(v_l)+\hat{min}(v_r)$. Thus the assertion holds in this case.
\end{proof}


\begin{Lemma} \label{lemma:D0_for_attach_D3_not_holds}
	If condition $D_3$ is false, there exists a vertex set $D\in\hat D_0(v)\cap\hat D_p(v)$ such that $D\cap\hat{TS}(v)\neq\emptyset$.
\end{Lemma}
\begin{proof}
	Because condition $D_3$ is not satisfied, we have either ``condition $D_2$ is not met'' or ``$\empts(v_l)\wedge\emptp(v_r)=1$''. First, consider the case when ``condition $D_2$ is not met''. Notice that, for the subcase when $\hat\alpha(v_r)>\hat\beta(v_l)$, we have exactly the same conditions as case $C_1$, which is mentioned in Subsection~\ref{subsection:attachment}. By combining the results of Lemmas~\ref{lemma:attach_g gamma_h(G_L) + gamma_h(G_R) = min {gamma_k(G_L) + gamma_k(G_R)}} and \ref{lemma:attach_g gamma_0(G)=min {gamma_k(G_L) + gamma_k(G_R)}}, we obtain that $\hat\gamma_0(v) = \min \setof{\hat\gamma_i(v_l)+\hat\gamma_i(v_r) \mid 1\le i \le min\text{-}ts}$. Thus, there exists a vertex set $D \in\hat D_0(v)$ such that $D\cap\hat V(v_l) \in\hat D_k(v_l)$ and $D\cap\hat V(v_r) \in\hat D_k(v_r)$ with $k>0$. Consequently, one can verify that $D\in\hat D_p(v)$ and $D\cap\hat{TS}(v)\neq\emptyset$. Similarly, for the subcase subcase when $\hat\alpha(v_r)\le \hat\beta(v_l)$, we have the same conditions as case $C_2$. And, one can show the assertion by applying Lemma~\ref{lemma:attach_s1 gamma_h(G_L) + gamma_h(G_R) = min {gamma_k(G_L) + gamma_k(G_R)}} and Lemma~\ref{lemma:attach_s1 gamma_0(G)=min {gamma_k(G_L) + gamma_k(G_R)}}. We omit the details here.
	
	Next, we consider the case when $\empts(v_l)\wedge\emptp(v_r)=1$ and assume that condition $D_2$ is true. According to Lemma~\ref{lemma:beta>0->empts=0}, we have $\hat\beta(v_l)=0$ as $\empts(v_l)=1$. Further, since the requirements of condition $D_2$ are met, we have $\hat\alpha(v_r)=\hat\beta(v_l)=0$. Combining  Lemma~\ref{lemma:determine_alpha_for_attach_s2} and Equation~(\ref{eq2}), we can get $\hat\alpha(v)=1$ and $\hat\gamma_0(v)=\hat\gamma_1(v)+1$. This implies that there exists a vertex set $D\in\hat D_0(v)\cap\hat D_p(v)$ such that $D\cap\hat{TS}(v)\neq\emptyset$. 
\end{proof}

In the following we establish the correctness of Procedure~\ref{algo:determine_empty_&_empts_for_attach} by Lemmas~\ref{lemma:determine_emptp_for_attach} and~\ref{lemma:determine_empts_for_attach}. Moreover, we continue to use the notations: $c_1, c_2, c_3$, and $c_4$, introduced in Subsection~\ref{subsection:emptp_empts_for_true}.

\begin{Lemma} \label{lemma:determine_emptp_for_attach}
	$
	\emptp(v) =
	\begin{cases}
		c_1\wedge c_2\wedge c_3\wedge c_4 & \text{if condition $D_3$ holds,}\\
		0   & \text{otherwise.}
	\end{cases}
	$
\end{Lemma}
\begin{proof}
	By Lemma~\ref{lemma:D0_for_attach_D3_not_holds}, if condition $D_3$ is false, then there exists a vertex set $D\in\hat D_0(v)\cap\hat D_p(v)$. This implies that $\emptp(v)=0$. Then, we consider the case when condition $D_3$ holds. Since $\hat G(v)=\hat G(\ddot v)$ and $\hat\gamma_0(v)=\hat\gamma_0(\ddot v)$ by Lemma~\ref{lemma:D0_for_attach_D3_holds}, one can verify that $D\in\hat D_0(v)\cap\hat D_p(v)$ if and only if $D\in\hat D_0(\ddot v)\cap\hat D_p(\ddot v)$. Combining Lemma~\ref{lemma:determine_emptp_for_true}, we have  $\emptp(v) = \emptp(\ddot v) = c_1\wedge c_2\wedge c_3\wedge c_4$. 
\end{proof}

\begin{Lemma} \label{lemma:determine_empts_for_attach}
	$
	\empts(v) =
	\begin{cases}
		\empts(v_l) & \text{if condition $D_3$ holds,}\\
		0   & \text{otherwise.}
	\end{cases}
	$
\end{Lemma}
\begin{proof}
	According to Lemma~\ref{lemma:D0_for_attach_D3_not_holds}, if condition $D_3$ is false, then there exists a vertex set $D\in\hat D_0(v)$ such that $D\cap\hat{TS}(v)\neq\emptyset$, which implies $\empts(v)=0$. Then, it remains to consider the case when condition $D_3$ holds. Notice that we have $\hat G(v)=\hat G(\ddot v)$ and $\hat\gamma_0(v)=\hat\gamma_0(\ddot v)$ by Lemma~\ref{lemma:D0_for_attach_D3_holds}.
	Therefore, one can see that $D\in\hat D_0(v)$ and $D\cap\hat{TS}(v) = \emptyset$ if and only if $D\in\hat D_0(\ddot v)$ and $D\cap\hat{TS}(v_l)= \emptyset$. Applying Lemma~\ref{lemma:D0_for_true_D2_holds}, we can get $D\in\hat D_0(\ddot v)$ if and only if $D_l\in\hat D_0(v_l)$ and $D_r\in\hat D_0(v_r)$. Thus, one can see that $D\in\hat D_0(\ddot v)$ and $D\cap\hat{TS}(v) = \emptyset$ if and only if $D_l\in\hat D_0(v_l)$ and $D_l\cap\hat{TS}(v_l)= \emptyset$. Consequently, $\empts(v)=\empts(v_l)$.
\end{proof}

\smallskip
Then we present the main result concerning  Procedure~\ref{algo:determine_empty_&_empts_for_attach}.

\begin{Lemma}\label{lemma:determine_emptp_empts_for_attach}
	Suppose that $v$ is annotated by attachment operation $\oplus$, then Procedure~\ref{algo:determine_empty_&_empts_for_attach} determines $\emptp(v)$ and $\empts(v)$ in $O(1)$ time.
\end{Lemma}
\begin{proof}
	Combining Lemmas~\ref{lemma:determine_emptp_for_attach} and~\ref{lemma:determine_empts_for_attach}, we obtain the correctness proof of the procedure. Moreover, the time complexity of the algorithm is $O(1)$ as all the steps can be easily implemented in $O(1)$ time.
\end{proof}

Finally, the following lemma, holding for attachment operation $\oplus$, is showed for completing the proof of Lemma~\ref{lemma:auxiliary_for_emptp_&_empts}. The proof is virtually identical to that given for Lemma~\ref{lemma:property1_holds_for_true}, and so we omit it.

\begin{Lemma}\label{lemma:property1_holds_for_attach}
	Suppose that $v$ is labeled by attachment operation $\oplus$. Then, we have $D\notin\hat D_p(v)$ or $D\cap\hat{TS}(v)=\emptyset$ for all vertex sets $D\in\hat D_0(v)$ if and only if $\emptp(v)\vee\empts(v)=1$. 
\end{Lemma}

\section{Conclusion and Future Work}\label{sec:conclusion}
This paper introduces an efficient algorithm for finding the minimum paired-dominating set on distance-hereditary graphs, significantly improving the time complexity from the previously known $O(n^2)$ to a faster $O(n+m)$. With a given decomposition tree of the graph, the execution time of our algorithm is further reduced to $O(n)$. Our approach employs dynamic programming to iteratively compute crucial values like $\hat\gamma_{p}(v)$, $\hat{min}(v)$, $\hat\alpha(v)$, $\hat\beta(v)$, $\empts(v)$, and $\emptp(v)$.
Moreover, a notable contribution of this research lies in establishing Equations (\ref{eq1}) and (\ref{eq2}), which describe pivotal mathematical relationships among these values, streamlining their determination process. These insights and algorithmic techniques enhance our ability to address the minimum paired-dominating set problem in distance-hereditary graphs.

The paired-dominating problem has attracted significant attention in recent years, yet several intriguing related questions remain unresolved. While distance-hereditary graphs constitute a subset of circle graphs, the complexity of the paired-dominating problem on circle graphs remains an open problem. Given that the classical domination problem and numerous of its variants are known to be NP-complete on circle graphs, a reasonable conjecture is that the paired-dominating problem also retains its NP-completeness.
Notably, Damian et al.~\cite{Damian02} introduced a $(3+\epsilon)$-approximation algorithm for the total domination problem on circle graphs. Given the technical correlations between these two problems, a natural extension would involve investigating the feasibility of a $(3+\epsilon)$-approximation algorithm for the paired-dominating problem.
Furthermore, in practical applications, planar graphs emerge naturally and find relevance in fields such as VLSI design systems, facility location problems, and security services. While the NP-completeness of the paired-domination problem has been established for planar graphs~\cite{Tripathi22}, an intriguing avenue for future research lies in the development of approximation algorithms tailored specifically for planar graphs. This could potentially open up new directions for tackling the paired-dominating problem in practical scenarios.

\section*{Declarations}
\subsubsection*{Competing Interests} The authors declare that they have no conflicts of interest, whether financial or non-financial, that could have influenced the work presented in this submission for publication.




\bibliographystyle{splncs04}
\bibliography{PDom2}

%
%
%
%
%

%
%
%
%
%
%
%
%
%
%
%
%

%
%
%
%

\bigbreak

\baselineskip 14pt


%
%
%
%

\end{document}